\documentclass[preprint2]{aastex6}
\usepackage{graphicx}
\usepackage{rotating}
\usepackage{amsmath}
\usepackage{natbib}
\usepackage{placeins}
\usepackage{aasref}
\usepackage{afterpage}

\begin{document}

\title{Bright clumps in the D68 ringlet near the end of the Cassini Mission}
\author{M.M. Hedman}
\affil{Physics Department, University of Idaho, Moscow ID 83844-0903}

\begin{abstract}
The D68 ringlet is the innermost narrow feature in Saturn's rings. Prior to 2014, the brightness of this ringlet did not vary much with longitude, but sometime in 2014 or 2015 a series of bright clumps appeared within D68. These clumps were up to four times brighter than the typical ringlet, occured within a span of $\sim120^\circ$ in corotating longitude, and moved at an average rate of 1751.7$^\circ$/day during the last year of the Cassini mission. The slow evolution and relative motions of these clumps suggest that they are composed of particles with a narrow (sub-kilometer) spread in semi-major axis. The clumps therefore probably consist of fine material released by collisions among larger (up to 20 meters wide) objects orbiting close to D68.  The event that triggered the formation of these bright clumps is still unclear, but it could have some connection to the material observed when the Cassini spacecraft passed between the planet and the rings.
\end{abstract}

\maketitle

\section{Introduction}

The D ring is the innermost component of Saturn's ring system, and it is a very complex region with structures on a broad range of scales. One of the more perplexing features in this region is a narrow ringlet found around 67,630 km from Saturn's center. This ringlet, designated D68, was first observed in a small number of images obtained by the Voyager spacecraft \citep{Showalter96}, and more recently has been imaged repeatedly by the cameras onboard the Cassini spacecraft, enabling several aspects of its structure and composition to be documented. These images show that the ringlet is very faint in back-scattered light, and that its brightness increases dramatically at higher phase angles \citep{HS15}. This implies that the visible material in this ringlet is very tenuous, and composed primarily of dust-sized particles in the 1-100 micron size range.  Meanwhile, high-resolution images show that D68 has a full-width at half-maximum of only around 10 km \citep{Hedman07}, while lower-resolution images reveal that D68 has a substantial orbital eccentricity ($ae\simeq$ 25 km)  and that its mean radial position appeared to oscillate $\pm10$ km around 67,627 km with a period of order 15 years \citep{Hedman14}.  Finally, these studies found that prior to 2014 the brightness of the ringlet had broad and subtle (roughly $\pm 25\%$) brightness variations that revolved around the planet at around 1751.65$^\circ$/day, consistent with the expected rate for material orbiting at the ringlet's observed mean radius \citep{Hedman14}.  

\begin{figure}
\centerline{\resizebox{3in}{!}{\includegraphics{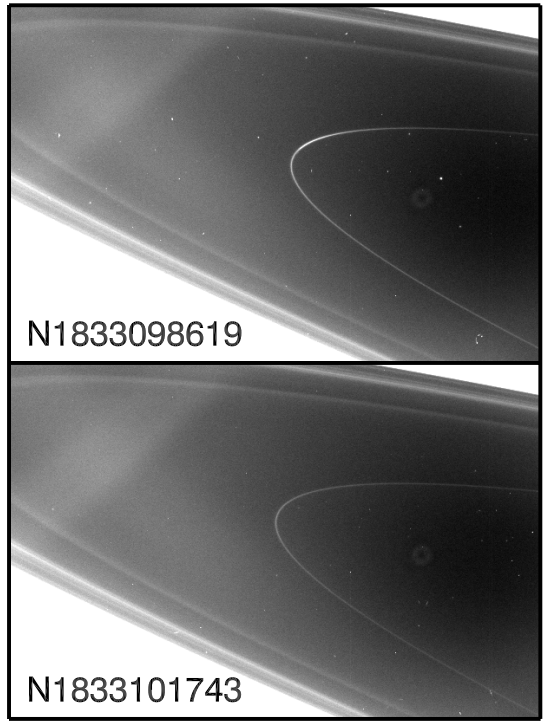}}}
\caption{Example of a clump in the D68 ringlet. Both of the above images were obtained on Day 33 of 2016 at phase angles around 126$^\circ$, and are shown with a common stretch. D68 is the narrow ring feature visible at the center of both images. The lower image corresponds to the typical appearance of D68, while the upper image shows a clear localized brightness enhancement near the ansa. Note the faint diagonal bands in the upper left of both panels are stray-light artifacts \citep{West10}.}
\label{clumpim}
\end{figure}

After 2014, the Cassini spacecraft continued to monitor D68 until the end of its mission in 2017. This was not only because D68 is scientifically interesting, but also because Cassini's final orbits around Saturn took it between the planet and the D ring, causing the spacecraft to pass within a few thousand kilometers of D68. Hence it was important to know how this ringlet was behaving in case it could either pose a hazard to the spacecraft or have any interesting effects on the in-situ measurements during Cassini's close encounters with Saturn. These images revealed unexpected and rather dramatic changes in the brightness structure of this ringlet. Whereas the brightness variations in D68 prior to 2014 could not be clearly discerned in individual images, images taken after 2015 showed a series of bright ``clumps'' that were several times brighter than the rest of the ringlet (see Figure~\ref{clumpim}).  These clumps were observed multiple times over the last two years of the Cassini mission, enabling their motion and slow evolution to be documented. 

Localized brightness enhancements have previously been observed in a number of other dusty rings. Some, like the arcs in Saturn's G ring and Neptune's Adams ring, persist for decades and therefore probably represent material actively confined by either mean-motion resonances or co-orbiting moons \citep{Hubbard86, Sicardy91, Porco91, NP02, Hedman07g, Hedman09, Renner14, Showalter17}. Others, like the bright features seen in the F ring and the dusty ringlets in the Encke Gap, are more transient and therefore probably consist of material released by collisions and/or concentrated by interparticle interactions \citep{Showalter98, Showalter04, BE02, French14, Murray18, FB97, Hedman13}. The relatively sudden appearance of the clumps in D68, as well as their evolution over the last two years of the Cassini mission, are more consistent with the latter scenario. Hence this work will explore the possibility that these clumps consist of material released by collisions among larger objects within D68.

The relevant aspects of the observational data used here are provided in Section~\ref{obs}, while Section~\ref{results} describes the properties of the D68 clumps, including their motions and brightness evolution. Section~\ref{discussion} then discusses how these features might have been generated from repeated collisions among objects orbiting within or close to D68. Finally, Section~\ref{prox} provides estimates of where these clumps were located relative to the Cassini spacecraft during its final orbits, and Section~\ref{summary} summarizes the results of this analysis.

\begin{sidewaystable*}
\vspace{4in}\caption{Summary of D-ring observations}
\label{dattab}
\resizebox{9in}{!}{\begin{tabular}{lc ccc ccc cc cc}\hline
Observation Name &$^a$ & Images$^b$ & \# & Observation Time & Duration & Inertial  & Phase  & Emission & Incidence& 
Radial & Noise  \\
& & && & & Longitude & Angle & Angle & Angle & Resolution & Level$^c$  \\ \hline
Rev 037 AZDKMRHP & A & N1547138243-N1547168273 &  73 & 2007-010T20:07:56 &   8.3 hours & 285.5$^\circ$-308.0$^\circ$ & 162.2$^\circ$-162.7$^\circ$ &  61.4$^\circ$- 64.1$^\circ$ & 104.2$^\circ$ & 
 10.2 km & 4.8 m \\
Rev 039 HIPHAMOVD & A & N1550157836-N1550176479 &  60 & 2007-045T17:26:01 &   5.2 hours & 288.2$^\circ$-309.0$^\circ$ & 161.9$^\circ$-162.2$^\circ$ &  62.7$^\circ$- 65.5$^\circ$ & 103.7$^\circ$  & 
8.8 km & 4.3 m \\
Rev 168 DRCLOSE  & A & N1719551308-N1719564388 &  25 & 2012-180T06:05:11 &   3.6 hours & 137.9$^\circ$-159.6$^\circ$ & 145.0$^\circ$-152.0$^\circ$ & 106.0$^\circ$-107.7$^\circ$ & 75.1$^\circ$  &
4.2 km &  6.1 m \\
Rev 173 DRNGMOV & A  & N1728999806-N1729023557 &  30 & 2012-289T16:08:05 &   6.6 hours & 304.9$^\circ$-324.9$^\circ$ & 141.5$^\circ$-145.4$^\circ$ & 124.6$^\circ$-125.8$^\circ$ & 73.8$^\circ$  &
8.5 km & 4.5 m \\
Rev 198 DRNGMOV  & B & N1761014449-N1761057029 &  52 & 2013-294T07:38:59 &  11.8 hours & 148.1$^\circ$-333.8$^\circ$ & 137.2$^\circ$-140.1$^\circ$ & 130.6$^\circ$-132.9$^\circ$ & 69.9$^\circ$  &
14.9 km & 11.8 m \\
Rev 201 DRNGMOV  & A  & N1770684650-N1770749910 &  40 & 2014-041T08:56:58 &  18.1 hours & 135.0$^\circ$-251.0$^\circ$ & 150.9$^\circ$-156.2$^\circ$ &  83.1$^\circ$- 88.4$^\circ$ & 68.8$^\circ$  &
11.9 km & 1.8 m \\
\hline
Rev 206 DRLPMOV & B  & N1784068515-N1784082291 &  17 & 2014-195T23:30:38 &   3.8 hours & 119.1$^\circ$-258.2$^\circ$ &  24.6$^\circ$- 26.7$^\circ$ &  92.0$^\circ$- 93.9$^\circ$ & 67.5$^\circ$  & 
9.0 km &  17.8 m  \\
Rev 211 DRCLOSE  & B & N1799468904-N1799472444 &   2 & 2015-009T03:57:09 &   1.0 hours &  -14.3$^\circ$-231.0$^\circ$ &  27.0$^\circ$- 29.0$^\circ$ &  93.0$^\circ$- 94.5$^\circ$ & 66.2$^\circ$  &
3.6 km & 10.8 m \\
Rev 212 DRCLOSE  & B & N1802349651-N1802351228 &   4 & 2015-042T11:52:53 &   0.4 hours & 325.5$^\circ$-343.7$^\circ$ & 139.5$^\circ$-140.2$^\circ$ & 101.5$^\circ$-101.8$^\circ$ & 66.0$^\circ$  & 
4.1 km & 0.9 m \\
Rev 213 DRCLOSE  & X & N1805090849-N1805092190 &   2 & 2015-074T05:17:19 &   0.4 hours & 310.1$^\circ$-339.2$^\circ$ & 126.0$^\circ$-127.2$^\circ$ &  96.4$^\circ$- 96.5$^\circ$ & 65.8$^\circ$  & 
3.4 km & ---\\
Rev 214 SATELLORB & X & N1807141302-N1807233113 &   2 & 2015-098T11:25:14 &  25.5 hours & 265.0$^\circ$-292.0$^\circ$ & 100.3$^\circ$-107.6$^\circ$ &  90.3$^\circ$- 90.3$^\circ$ & 65.6$^\circ$  &
9.4 km & --- \\
Rev 218 HPLELR  & B  & N1815051394-N1815058482 &   5 & 2015-189T12:57:20 &   2.0 hours & 205.0$^\circ$-240.0$^\circ$ & 148.6$^\circ$-149.0$^\circ$ &  90.4$^\circ$- 90.4$^\circ$ & 65.1$^\circ$  &
10.1 km & 2.2 m \\
Rev 222 HPLELR  & X & N1822499001-N1822499041 &   2 & 2015-275T16:40:29 &   0.0 hours & 210.0$^\circ$-270.0$^\circ$ & 136.8$^\circ$-136.8$^\circ$ &  90.6$^\circ$- 90.6$^\circ$ & 64.7$^\circ$  &
7.7 km & --- \\
Rev 227 MUTUALEVE & X & N1827866095-N1827866251 &   2 & 2015-337T19:32:28 &   0.0 hours & 270.0$^\circ$-300.0$^\circ$ & 101.4$^\circ$-101.4$^\circ$ &  89.5$^\circ$- 89.5$^\circ$ & 64.4$^\circ$  &
27.9 km & --- \\
Rev 228 SATELLORB & X & N1828944662-N1828944662 &   1 & 2015-350T07:07:09 &   0.0 hours & 270.0$^\circ$-300.0$^\circ$ & 102.8$^\circ$-102.8$^\circ$ &  89.5$^\circ$- 89.5$^\circ$ & 64.4$^\circ$  &
31.5 km & --- \\
Rev 228 HPLELR  & B  & N1829382845-N1829400801 &  68 & 2015-355T11:19:45 &   5.0 hours & 225.0$^\circ$-255.0$^\circ$ & 138.0$^\circ$-139.9$^\circ$ &  90.4$^\circ$- 90.5$^\circ$ & 64.3$^\circ$  & 
7.4 km & 2.5m \\
Rev 231 DRCLOSE & B  & N1832861779-N1832868799 &   6 & 2016-030T16:10:32 &   1.9 hours &  10.9$^\circ$-222.3$^\circ$ & 154.7$^\circ$-157.7$^\circ$ &  93.2$^\circ$- 93.5$^\circ$ & 64.2$^\circ$  &  
2.9 km & 5.6 m \\
Rev 231 FNTHPMOV & B  & N1833090809-N1833111115 &  14 & 2016-033T09:38:23 &   5.6 hours & 195.6$^\circ$-311.8$^\circ$ & 125.9$^\circ$-127.6$^\circ$ &  86.3$^\circ$- 86.8$^\circ$ & 64.2$^\circ$  &
9.9 km & 5.8 m \\
\hline
Rev 233 FNTHPMOV & B & N1836371570-N1836388706 &  57 & 2016-071T08:30:58 &   4.8 hours &  45.0$^\circ$- 60.0$^\circ$ & 143.5$^\circ$-148.0$^\circ$ &  93.7$^\circ$- 95.6$^\circ$ & 
64.0$^\circ$  &2.1 km & 2.9 m \\
Rev 245 HPCLOSE & A  & N1855239471-N1855259479 &  48 & 2016-289T17:57:50 &   5.6 hours & 225.0$^\circ$-250.0$^\circ$ & 153.5$^\circ$-158.1$^\circ$ &  96.7$^\circ$-100.4$^\circ$ & 63.5$^\circ$  &
6.5 km & 1.6 m \\
Rev 256 HPMONITOR & A & N1862795920-N1862813869 &  29 & 2017-011T04:40:42 &   5.0 hours & 220.0$^\circ$-250.0$^\circ$ & 156.1$^\circ$-159.1$^\circ$ &  96.9$^\circ$- 99.5$^\circ$ & 63.3$^\circ$  & 
6.3 km & 2.3 m \\
Rev 263 HIPHASE  & A & W1867048670-W1867048670 &   1 & 2017-060T07:29:52 &   0.0 hours & 190.9$^\circ$-248.3$^\circ$ & 174.5$^\circ$-174.5$^\circ$ & 119.6$^\circ$-119.6$^\circ$ & 63.3$^\circ$  &
37.2 km & 2.4 m \\
Rev 272 HIPHASE & A  & W1872474248-W1872492064 &  35 & 2017-123T05:04:03 &   4.9 hours & 195.0$^\circ$-245.0$^\circ$ & 163.5$^\circ$-171.3$^\circ$ & 111.7$^\circ$-120.9$^\circ$ & 63.3$^\circ$  &
27.8 km & 2.2 m \\
Rev 279 HPMONITOR & A & N1876589648-N1876613228 &  16 & 2017-170T21:01:39 &   6.5 hours & 210.0$^\circ$-240.0$^\circ$ & 148.5$^\circ$-150.1$^\circ$ &  85.3$^\circ$- 86.9$^\circ$ & 63.3$^\circ$  &
7.6 km & 2.2 m \\
Rev 289 HPMONITOR & A & N1881697721-N1881715322 &  60 & 2017-229T23:05:45 &   4.9 hours &  20.0$^\circ$- 71.0$^\circ$ & 142.8$^\circ$-144.0$^\circ$ &  79.5$^\circ$- 80.7$^\circ$ & 63.3$^\circ$  & 
7.5 km & 2.9 m\\
\hline
\end{tabular}}
$^a$ Regions used for integration and background subtraction: A: Integrate over 67,500-67,750 km, background computed from regions 66,500-67,500 km and  67,750-68,750 km, B: Integrate ove 67,550-67,750 km and background computed from regions 67,500-67,550 and 67,750-67,800 km, X= none (not plotted).

$^b$ Images names start with the letter N if taken with the Narrow Angle Camera and W if taken with the Wide Angle Camera, and the following numbers correspond to the spacecraft clock time when the image was taken.

$^c$ Rough estimate of the instrumental noise in the PC-NEW profiles. The reported numbers are the standard deviations of the data values sampled at 0.1$^\circ$, after applying a 3$^\circ$-wide high-pass filter to remove features like the clumps. 
\end{sidewaystable*}

\section{Cassini ISS observations}
\label{obs}

The data on the D68 ringlet considered here come from the Imaging Science Subsystem (ISS) onboard the Cassini Spacecraft \citep{Porco04, West10}. Table~\ref{dattab} summarizes the images used  in this analysis. All of these images were calibrated using the standard CISSCAL routines that remove dark currents, apply flatfield corrections, and convert the observed brightness data to $I/F$, a standardized measure of reflectance that is unity for a Lambertian surface illuminated and  viewed at normal incidence \citep{Porco04}. These calibrated images were geometrically navigated with the appropriate SPICE kernels, and the pointing was refined as needed based on the observed locations of stars in the field of view. Note that the long exposure durations used for many images (typically 1-20 seconds) caused the images of stars to be smeared into streaks. The algorithms for navigating images based on star streaks are described in \citet{Hedman14}. 

In previous analyses of D68 images, the brightness data from each image would be averaged over longitude to produce a radial brightness profile. This was sensible when the ringlet showed only weak longitudinal brightness variations, but is no longer appropriate now that the ringlet possesses clumps that are smaller than the longitude range spanned by a single image (see Figure~\ref{clumpim}). Hence, for this analysis the image data were instead re-projected onto regular grids in radii $r$ and inertial longitudes $\lambda_i$\footnote{Inertial longitude is measured relative to the ascending node of the rings in the J2000 coordinate system}. Each column of the re-projected maps then provides a radial profile of D68 at a single inertial longitude, which can be co-added as needed to generate longitudinal profiles with sufficient resolution to document the clumps. Since the ring material orbits the planet, these profiles are constructed in a co-rotating longitude system $\lambda_c=\lambda_i-n_0(t-t_0)$ where $n_0$ is the mean motion of the ring material, $t$ is the observation time, and $t_0$ is a reference time. This study uses a reference time of  300000000 TDB (seconds past J2000 epoch) or 2009-185T17:18:54 UTC, which is the same value used in prior investigations of this ringlet's structure \citep{Hedman14}. Also, the mean motion is taken to be 1751.7$^\circ$/day, a value that ensures the most prominent clump remains at nearly the same co-rotating longitude in the available data. This rate is also consistent with the expected mean motion of particles orbiting within D68. Note that material moving at this rate will smear over 0.02$^\circ$-0.4$^\circ$ in longitude over the 1-20 second exposure times of the relevant images. Fortunately, this longitudinal smear is small compared to the scale of the clumps that form the focus of this study.

Cassini images rarely have sufficient spatial resolution to discern any of D68's radial structure\footnote{Substructure within D68 is only seen in images with resolutions better than 3 kilometers per pixel \citep{Hedman07},  while most of the useful images listed in Table~\ref{dattab} have coarser resolution.}, so for this study the ringlet's brightness is quantified in terms of its equivalent width (EW), which is the radially integrated $I/F$ of the ringlet above the background:

\begin{equation}
EW = \int (I/F-I/F_{\rm back}) dr
\end{equation}
In practice, this integral is performed over a limited range of the radii, and the background signal is given by a linear fit to the signal on either side of these regions. The exact ranges used for these calculations depend on the resolution and the field of view of the relevant images. Whenever possible, the integral was performed  over the radius range of 67,500-67,750 km, and the background was computed from the data in  regions 1000-km wide on either side of this range. These choices are consistent with those used for many of the images examined by \citet{HS15}, and the  central range does encapsulate the entirety of D68 in all of the images used for this analysis. However, for certain observations (see Table~\ref{dattab}), D68 fell close to the edge of the camera's field of view, and so using such extensive radial ranges was not feasible (Note that
 all of the observations from around the time when the clumps likely formed have this property, so completely excluding these observations would be inappropriate). In these cases, the integral was instead performed over the slightly restricted radius range of  67,550-67,750 km, and the background signal was determined from  regions only 50-km wide on either side of this central range. For all of these sequences,  D68 was fully inside the reduced range, and tests showed that changing the ranges did not substantially affect the EW estimates (typical changes being less than 10\%). 

Of course, the observed brightness of D68 not only depends on the amount of material in the ringlet, but also on the viewing geometry, which is parameterized by the incidence, emission and phase angles. Fortunately, in this case the dependence on incidence and emission angles are relatively simple because D68 has a very low optical depth. While D68's optical depth has not yet been directly measured because it has so far eluded detection in occultations, the overall brightness of the ringlet is consistent with the peak optical depth of the visible material being of order 0.001. This means that any individual particle is unlikely to either cast a shadow on or block the light from any other particle \citep{Hedman07, HS15}. In this limit, the surface brightness is independent of the incidence angle and is proportional to $1/|\mu|$, where $\mu$ is the cosine of the emission angle\footnote{Here emission angle is defined to be 0$^\circ$ when the ring is viewed from a point directly north of the rings, and 180$^\circ$ when viewed from directly south of the rings.} Hence  the above estimates of the ringlet's equivalent width (EW) are multiplied by $|\mu|$ to obtain the so-called normal equivalent width (NEW)\footnote{This notation is consistent with that used in previous photometric studies of the D ring \citep{Showalter96, Hedman07}. Note that occultation studies of narrow rings designate this quantity as simply the Equivalent Width \citep[see, e.g.][]{French91}}. 

The variations in  D68's brightness with phase angle are more complex, but fortunately Cassini observations obtained prior to the clump's formation have yielded detailed measurements of  D68's phase curve \citep{HS15}. These observations show that D68 is strongly forward-scattering, and that the NEW of D68 varies by over two orders of magnitude across the range of phase angles listed in Table~\ref{dattab}. \citet{HS15} parameterized D68's observed phase function a number of ways, but for the sake of simplicity this analysis will use the analytical approximation of D68's phase curve consisting of three Henyey-Greenstein functions:

\begin{equation}
P(\theta)=\sum_{i=1}^3 \frac{w_i}{4\pi}\frac{1-g_i^2}{(1+g_i^2-2g_i\cos\theta)^{3/2}}
\label{phase}
\end{equation}
where $\theta$ is the scattering angle (i.e.\ the supplement of the phase angle), and the parameters are $g_1=0.995$, $g_2=0.585$ and $g_3=0.005$, with weights $w_1=0.754$, $w_2=0.151$ and $w_3=0.095$. Dividing the NEW values by this phase function yields a parameter we call the Phase-Corrected Normal Equivalent Width, or (PC-NEW), which is proportional to the ringlet's integrated normal optical depth (also known as Equivalent Depth). Note that this procedure assumes that the bright clumps have the same phase function from the previously quiescent ring. Fortunately,  this assumption appears to be justified since the above correction yields reasonably consistent PC-NEW values for both the clumps and the background ringlet. While there are some variations in D68's typical brightness after this correction, they are sufficiently small that they do not significantly hinder the following analysis. Indeed, this correction greatly facilitates comparisons among the observations.

\begin{figure*}[!p]
\resizebox{6.5in}{!}{\includegraphics{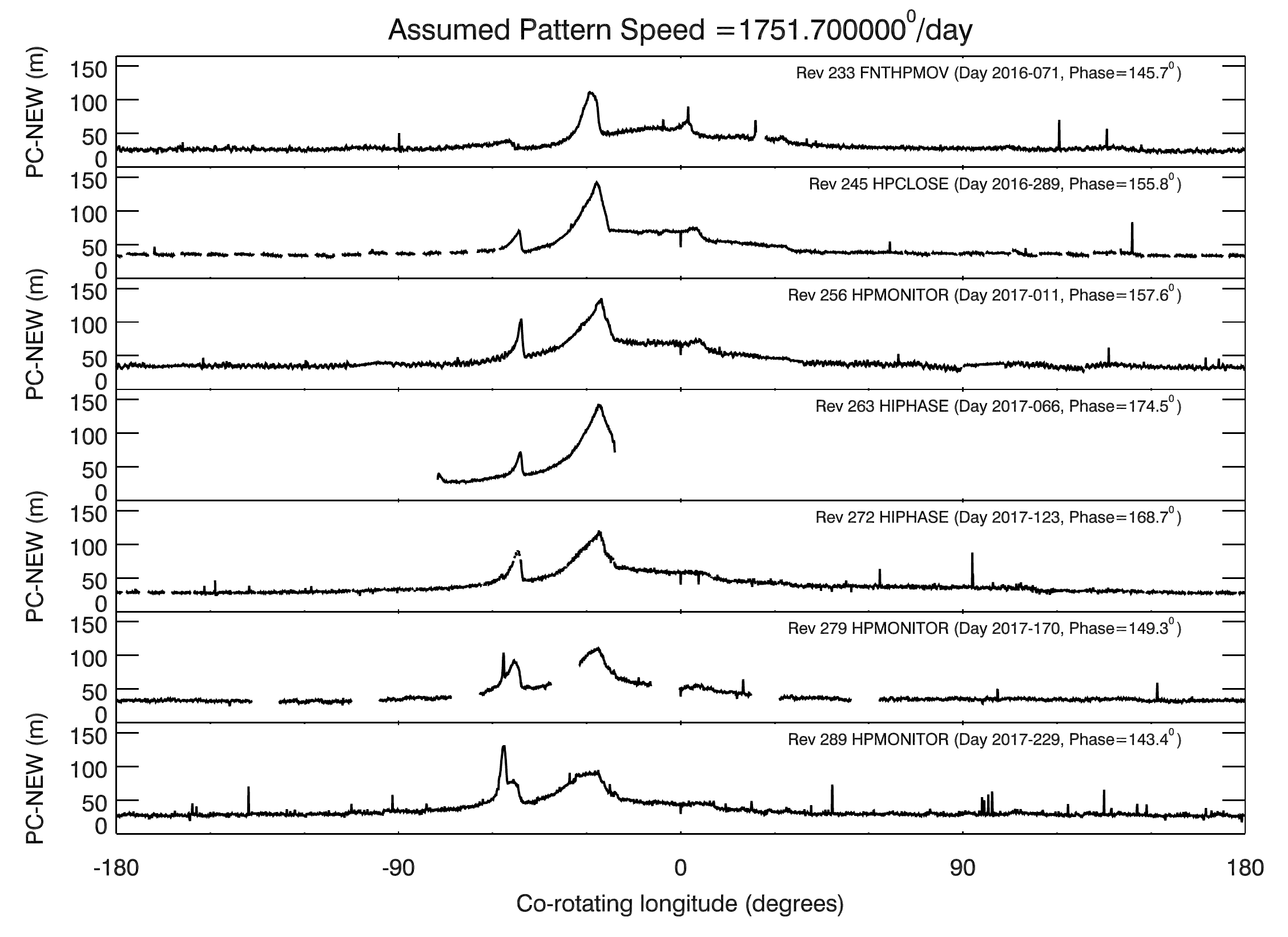}}
\caption{Longitudinal brightness profiles of D68 obtained in 2016-2017. Each panel uses data from one of the sequences listed in Table~\ref{dattab}, with individual brightness estimates averaged over co-rotating longitude bins 0.1$^\circ$ wide. The longitude system used here rotates at a speed of 1751.7$^\circ$/day and has an epoch time of 300000000 TDB (UTC 2009-185T17:18:54). Note that narrow spikes and dips correspond to instrumental artifacts like hot pixels, cosmic rays or background stars. All these profiles clearly show a series of peaks between $-75^\circ$ and $+45^\circ$  whose shape and position slowly evolve over time.}
\label{fullclump}
\end{figure*}

The profiles of D68's phase-corrected normal equivalent width versus co-rotating longitude are shown in Figures~\ref{fullclump}, ~\ref{preclump} and ~\ref{clumpform} and are provided in three supplemental tables to this work. Note that no error bars are provided on these profiles because systematic uncertainties due to phenomena like variations in the background level dominate over statistical uncertainties, and such systematic uncertainties are difficult to calculate {\it a priori}. Instead, rough estimates of these errors are computed based on the $rms$ variations in the profiles after applying a $3^\circ$-wide high-pass filter to suppress features like the clumps (the numerical values of these noise estimates do not depend strongly on the scale of the filter). These noise estimates are provided in Table~\ref{dattab}.

\nocite{fullclump2}

\section{Observable properties of the D68 Clumps}
\label{results}

This section summarizes the observable properties of the D68 clumps derived from the above profiles of the ringlet's brightness. Section~\ref{clumps} describes the distribution and evolution of the bright clumps observed in the last 18 months of the Cassini mission. Section~\ref{form} compares these clumps with the previously-observed brightness variations in the ring and uses the sporadic observations in 2014 and 2015 to constrain when these bright clumps may have originally formed. Finally, Section~\ref{evolution} documents the trends in the locations and brightnesses of these clumps.

\begin{figure*}
\resizebox{6.5in}{!}{\includegraphics{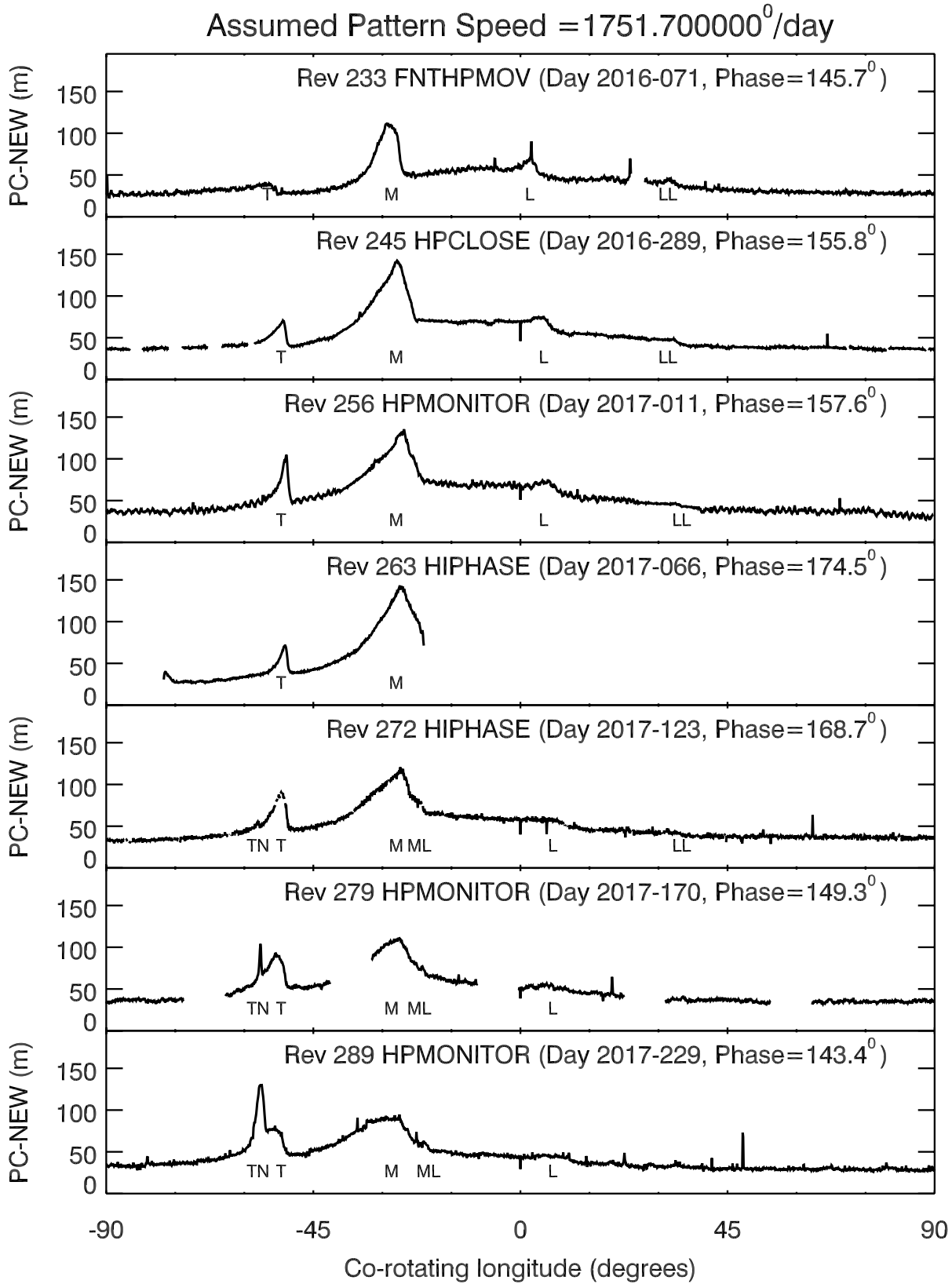}}
\caption{A closer look at ongitudinal brightness profiles of D68 obtained in 2016-2017. These are the same profiles shown in Figure~\ref{fullclump}, but zoomed in on the clump-rich region and with specific clumps labeled.}
\label{fullclump2}
\end{figure*}

\begin{figure*}
\resizebox{6.5in}{!}{\includegraphics{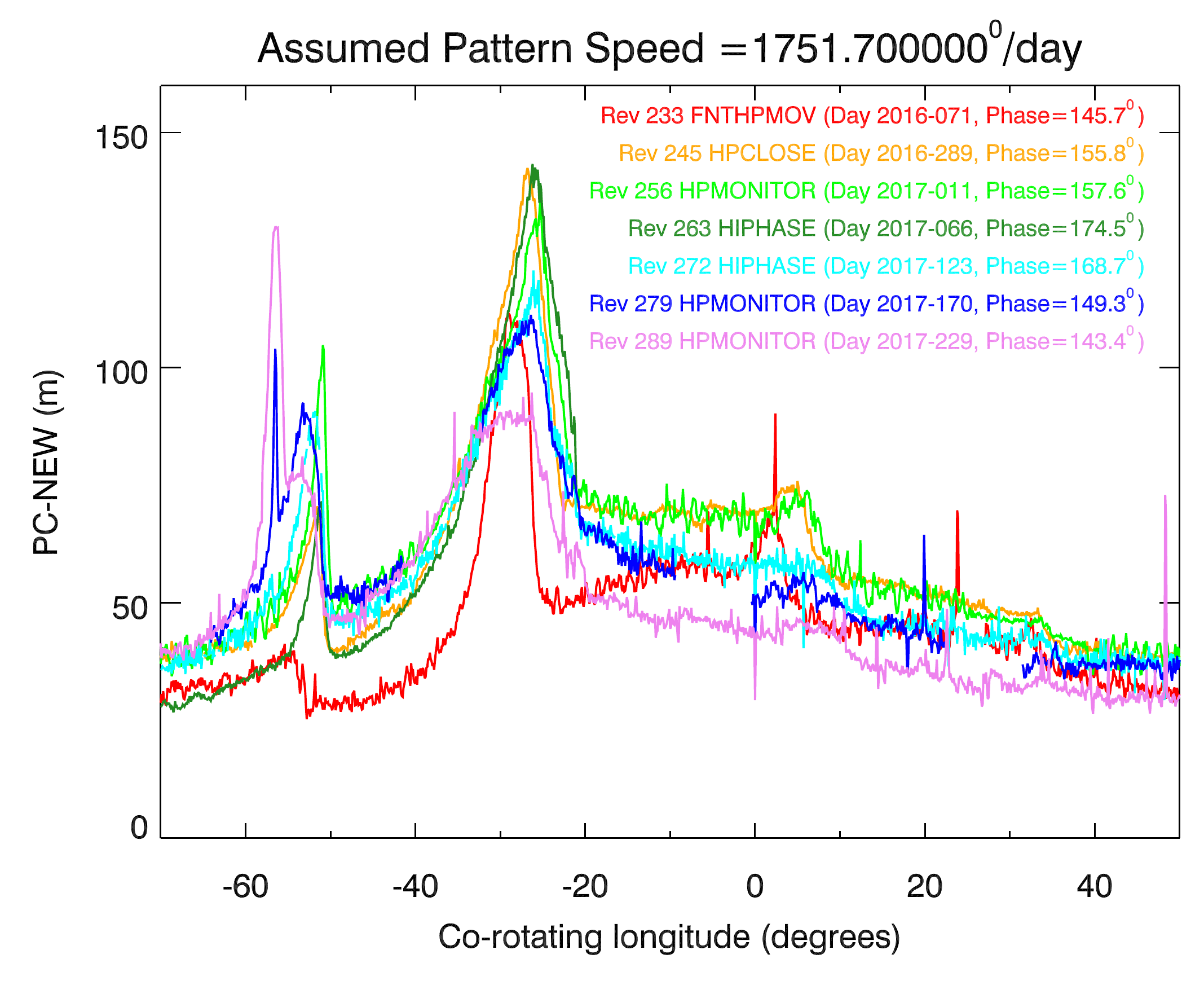}}
\caption{Another look at the longitudinal brightness profiles of D68 obtained in 2016-2017. These are the same profiles shown in Figures~\ref{fullclump} and~\ref{fullclump2}, but now are overlaid on top of each other to highlight changes in brightness and morphology of the clumps..}
\label{fullclump2x}
\end{figure*}

\subsection{The clumps during the final 18 months of the Cassini Mission}
\label{clumps}

Longitudinal brightness variations can be seen in every observation of D68 obtained in 2016 and 2017. Figure~\ref{fullclump} shows profiles derived from a sub-set of those observations that covered most of the clumps, were obtained at phase angles above 140$^\circ$ and had ring opening angles above 5$^\circ$.  These seven profiles provide the clearest picture of D68's structure during this time.  The estimated noise levels for all these profiles are less than 3 meters (see Table~\ref{dattab}), which is consistent with their generally smooth appearance outside of  a few sharp excursions that can be attributed to background stars and cosmic rays. Hence statistical noise and most systematic variations associated with instrumental phenomena like stray-light artifacts are less than 10\% of the signal for all of these profiles. Observations at lower phase and/or ring opening angles also captured these brightness variations, but either had lower signal-to-noise or did not provide such clean profiles due to the greatly degraded radial resolution away from the ansa.

In all seven of these profiles the bright clumps are clearly restricted to a range of longitudes between $\pm90^\circ$. Figures~\ref{fullclump2} and~\ref{fullclump2x} provide closer looks at these clumps and more clearly shows how they evolved over the last two years of the Cassini mission. In early 2016, there are four clear peaks in the ringlet's brightness. These four features are here designated with the letters T (for ``Trailing"), M (for ``Middle"), L (for ``Leading") and LL (for ``Leading Leading"). The M clump is the brightest of these features,  being 4-5 times brighter than the background ring, while the T, L and LL clumps are more subtle features. All four clumps appear to be superimposed on a broad brightness maximum that is centered between the M and L clumps. 

Over the course of 2016, each of these clumps evolved significantly. The L and LL clumps became progressively broader and less distinct. By contrast, the M clump became somewhat brighter and slightly more sharply peaked, while the T clump became much brighter and developed a strongly asymmetric shape with a very sharp leading edge.  In 2017, the L and LL clumps continued to become less and less distinct, with the LL clump becoming practically invisible by the latter half of 2017. The T and M clumps also started to become progressively broader over the course of 2017. At the same time, two new features appear in the profiles. First, a small bright feature, designated ML (for ``Middle Leading") emerges from the leading edge of the M clump and drifts ahead of the clump over the course of 2017. More dramatically, a new brightness maximum appears behind the T clump. Designated TN (for ``Trailing New"), this feature is first seen on Day 123 of 2017, where it appears as very faint peak on the trailing flank of the T clump. On Day 170, it appears as a narrow feature with a peak brightness intermediate between the T and M clumps. Finally, on Day 229 this clump has brightened and broadened dramatically, becoming the brightest feature in the ringlet.

\begin{figure*}[!p]
\resizebox{6.5in}{!}{\includegraphics{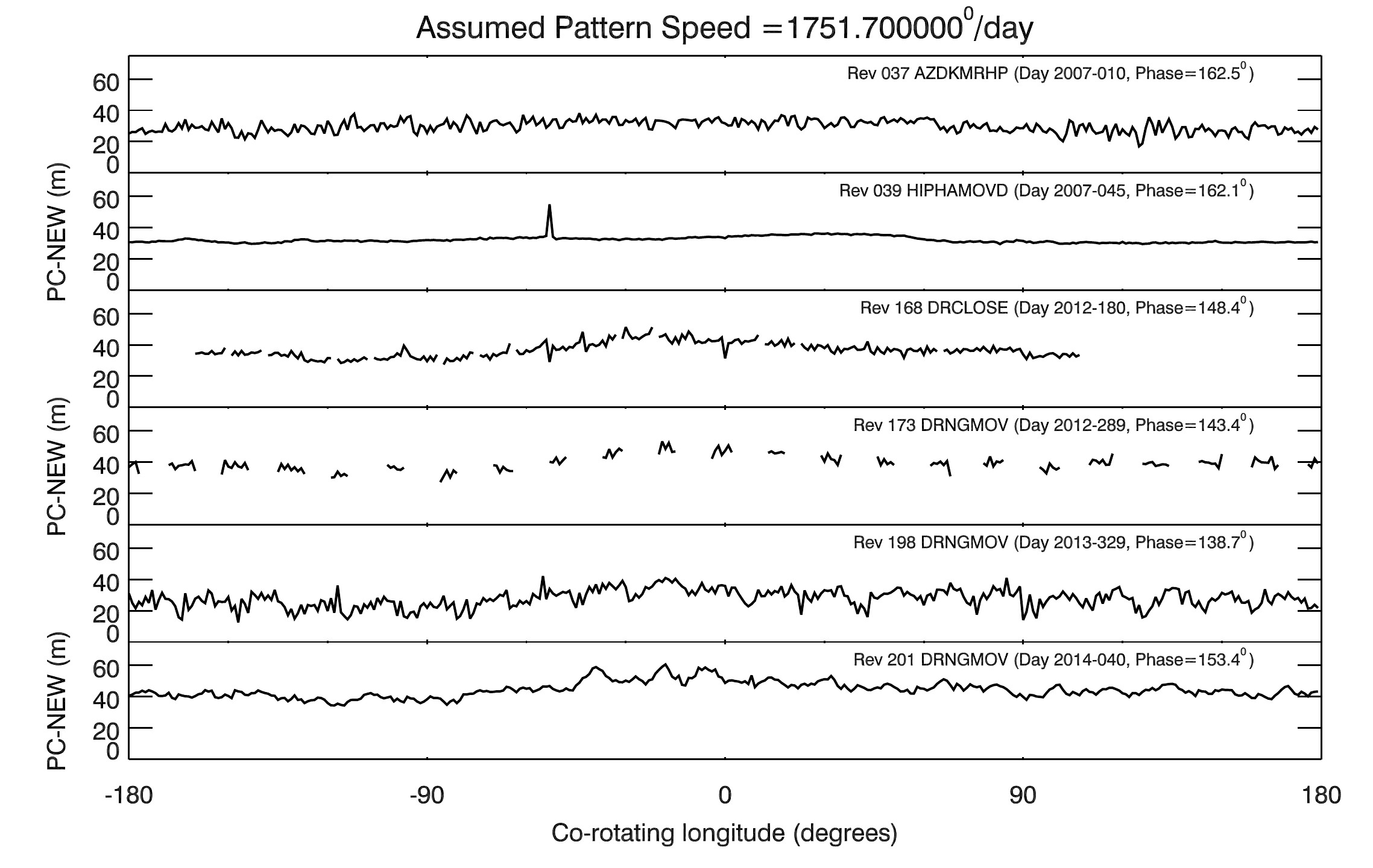}}
\caption{Longitudinal brightness profiles of D68 obtained through early 2014. Each panel uses data from one of the sequences listed in Table~\ref{dattab}, with individual brightness estimates averaged over co-rotating longitude bins 1$^\circ$ wide. The longitude system used here rotates at a speed of 1751.7$^\circ$/day and has an epoch time of 300000000 TDB (UTC 2009-185T17:18:54). Note the narrow spike in the Rev 039 HIPHAMOVD profile is an image artifact, as are the regular ripples in the Rev 198 DRNGMOV profile.
}
\label{preclump}
\end{figure*}

\begin{figure*}
\resizebox{6.5in}{!}{\includegraphics{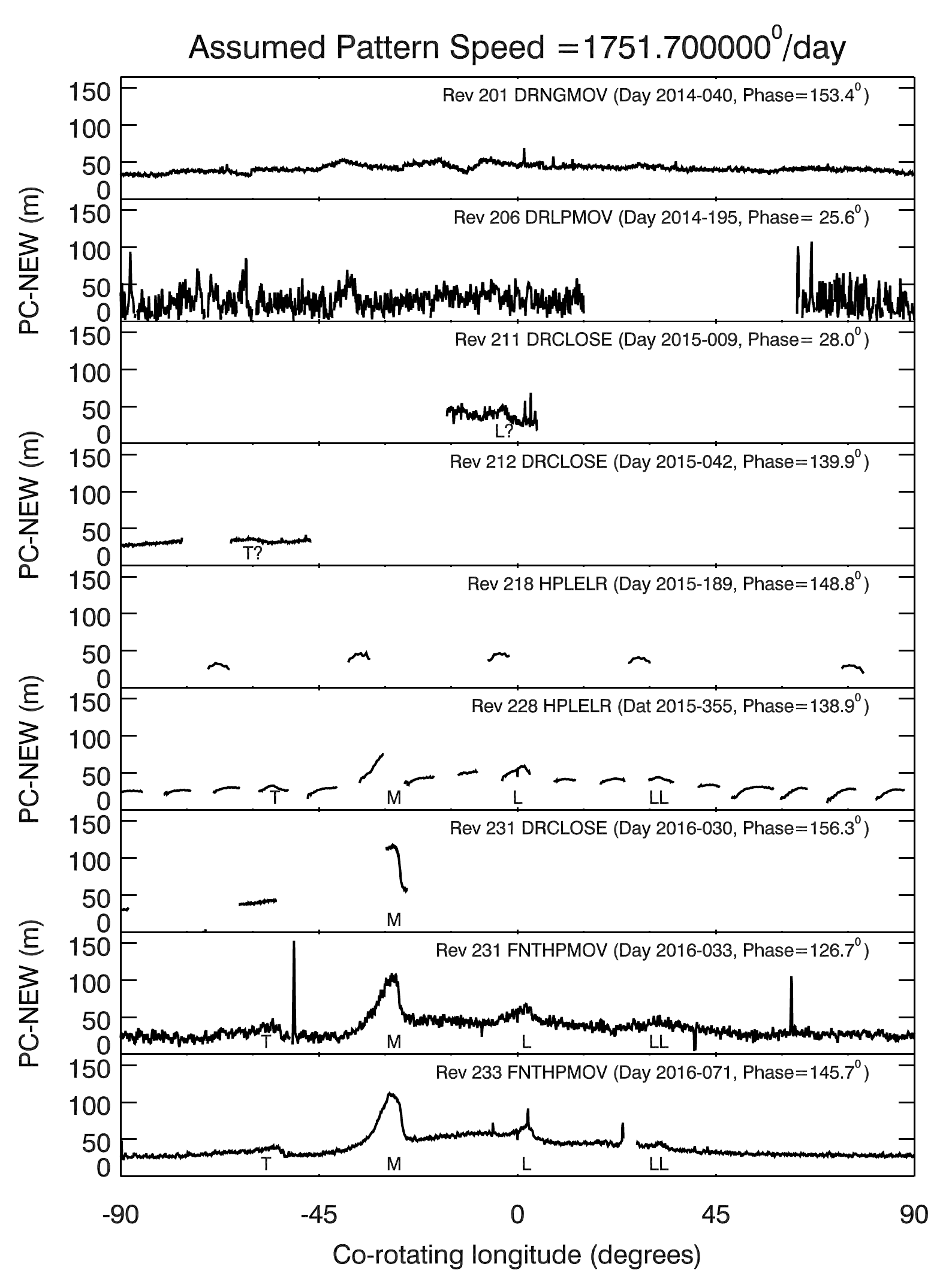}}
\caption{Longitudinal brightness profiles of D68 obtained between 2014 and 2016. Each panel uses data from one of the sequences listed in Table~\ref{dattab}, with individual brightness estimates averaged over co-rotating longitude bins 0.1$^\circ$ wide. The longitude system used here rotates at a speed of 1751.7$^\circ$/day and has an epoch time of 300000000 TDB (UTC 2009-185T17:18:54).}
\label{clumpform}
\end{figure*}

\subsection{Connections to earlier brightness variations and the origins of the clumps}
\label{form}

The bright clumps  described above are completely different from the brightness variations seen in D68 prior to 2014. Figure~\ref{preclump} shows profiles of the rings derived from six observations obtained prior to early 2014. Five of these are a subset of observations from \citet{Hedman14}, but have been processed using the techniques described above to ensure that any narrow structures would not be missed. Also note that the brightness profiles presented in \citet{Hedman14} used a slightly different co-rotating longitude system with a rotation rate of 1751.65$^\circ$/day, while here a rate of 1751.70$^\circ$/day is used to facilitate comparisons with the later data. Since there were no obvious sharp features in these profiles, these profiles were made with larger longitudinal bins (1$^\circ$ instead of 0.1$^\circ$) in order to improve signal-to-noise.

None of these early observations show any of the bright clumps seen in 2016 and 2017. Instead, any real brightness variations are much more subtle and can be comparable to variations associated with instrumental noise. Note that these profiles are more heterogeneous in their noise properties than those illustrated in Figure~\ref{fullclump} because they were obtained under a broader range of viewing geometries and employed a larger range of exposure times.

The Rev 039 HIPHAMOVD, Rev 168 DRCLOSE and Rev 173 DRNGMOV profiles are comparable in quality to the 2016-2017 profiles, exhibiting  random fine-scale variations of around 5 m.  The Rev 037 AZDKMRHP profile shows somewhat higher scatter in its profile, despite having a similar noise level in Table~\ref{dattab}. This is because the variations seen here are primarily at the few-degree scale due to instrumental background phenomena like stray light artifacts \citep{West10}.  Still, all four of these early profiles show the broad, low peak that was first identified in \citet{Hedman14} and appears to be a real ring structure. In 2007, the most visible aspect of this modulation is a falling slope in brightness around +90$^\circ$ in the new co-rotating coordinate system. Between 2007 and 2012, the shape and location of the peak shifted, such that in 2012-2013 the most obvious brightness variation is now a rising slope centered around $-45^\circ$, near the location where the clumps would later appear. This same brightness maximum is also present in the Rev 198 DRNGMOV profile, but it is somewhat harder to discern because this profile shows a periodic modulation with a wavelength of order 20$^\circ$, which arises because there is a particularly bright stray-light artifact running across part of these images that is not entirely removed by the background-subtraction procedures.

The Rev 201 DRNGMOV profile is interesting because on fine scales it appears to be of comparable quality to the Rev 039 HIPHAMOVD and Rev 168 DRCLOSE profiles, but it also exhibits some novel brightness structure. Specifically, in addition to the broad hump, this profile also has brightness variations on scales of a few tens of degrees between $-45^\circ$ and $0^\circ$ longitude, where some of the brightest clumps would later be found. Stray light artifacts are far less prominent in this dataset than in Rev 198 DRNGMOV, so these could be real features in the ring. Unfortunately, there are no other profiles of comparably good quality from this time period that can be used to confirm the existence of these weak peaks, and so all that can be said with confidence at this time is that nothing like the bright clumps seen in 2016 were present in early 2014.\footnote{It is also potentially interesting that the phase-corrected brightnesses of the profiles obtained in 2012-2014 are roughly 30\% higher than those of the two profiles from 2007. Determining whether this is a real change in the ringlet or an error  in the assumed phase function will require a more detailed photometric analysis that is beyond the scope of this work.}

Observations of D68 were very limited between early 2014 and early 2016, in part because the spacecraft was close to Saturn's equatorial plane for part of that time. Table~\ref{dattab} includes a list of {\sl every} observation of D68 with resolution better than 30 km/pixel during this time period, and Figure~\ref{clumpform} shows longitudinal brightness profiles for those image sequences which captured more than 10$^\circ$ of the region between co-rotating longitudes of $\pm90^\circ$. These observations, which generally have lower signal-to-noise or less longitudinal coverage than those discussed above, still provide slightly improved constraints on when the clumps might have formed.

Working backwards from when the clumps clearly existed, we can first note that the T, M, L and probably LL clumps can all be seen in the Rev 231 FNTHPMOV observation obtained in early 2016, albeit at lower signal-to-noise. Portions of the M clump are also clearly visible in the Rev 231 DRCLOSE and the Rev 228 HPLELR obesrvations. The latter was obtained at very low ring opening angles, and therefore only provides limited snapshots of the ring's brightness. Also, these snapshots often show noticeable trends within the data derived from each image (e.g. the slopes seen beyond $+45^\circ$), which likely arise because the extreme foreshortening away from the ansa can lead to slight inaccuracies when the observed brightness values are interpolated onto maps of brightness versus radius and longitude.  Despite these complications, there are hints of the T, L and LL clumps in these data. The T, M, L and LL clumps therefore all probably existed by the end of 2015.

Unfortunately, there are no observations of the longitudes that would contain the M clump between mid-2014 and late 2015. One observation from mid-2015 is another low-ring-opening angle observation which just missed all four clumps. Prior to this, in early 2015, there are two  DRCLOSE observations. The Rev 211 DRCLOSE observation on Day 9 of 2015 shows a brightness maximum that could be the L  clump, while the Rev 212 DRCLOSE observation  on Day 42 shows a small brightness variation that could be the T clump.

Finally, in mid-2014 there was a low-phase DRLPMOV observation in Rev 206. The signal-to-noise for this observation is quite low because it was obtained at low phase angles where the ring is comparatively faint \citep{HS15}, and so it is hazardous to interpret any of the brightness variations in this profile as evidence for any of the later clumps, Still, it is worth noting that there are no features that are as bright as the M clump.  It is therefore reasonable to conclude that the bright M clump most likely formed sometime after mid-2014 and before late 2015, but that the fainter L and T clumps probably began to appear sometime in  2014.

\begin{table*}[!th]
\caption{Locations of selected D68 clump features in a co-rotating reference frame with a mean motion of 1751.7$^\circ$/day and an epoch time of 300000000 TDB}
\label{d68locs}
\hspace{-1in}\resizebox{7.5in}{!}{\begin{tabular}{l c c c c c c c c c c}\hline
Observation & Date & TN peak & T peak & T edge & M peak & * & ML peak & M/ML edge & L peak & LL peak \\ \hline
Rev 211 DRCLOSE &
2015-009    &   --- &    --- &    --- &    --- &    --- &    --- &    --- &    -3.5$^\circ$ &    --- \\
Rev 212 DRCLOSE &
2015-042    &  --- &   -60.1$^\circ$ &   -56.5$^\circ$ &  ---  &  --- &   --- &  --- &   --- &   --- \\
Rev 228 HPLELR &
2015-355 &  --- &    -55.7$^\circ$   & --- &   ---- &    --- &    --- &    --- &    0.8$^\circ$  &  31.7$^\circ$ \\
Rev 231 FNTHPMOV &
2016-033   &    --- &    --- &    --- &    -28.2$^\circ$ &   --- &    --- &    -26.5$^\circ$ &  1.9$^\circ$     & 32.3$^\circ$ \\
Rev 233 FNTHPMOV &
2016-071   &    --- &    -55.4$^\circ$ &  -52.8$^\circ$ &   -29.0$^\circ$  & --- &    --- &    -26.2$^\circ$ &   2.3$^\circ$ &     32.5$^\circ$ \\
Rev 245 HPCLOSE &
2016-289   &    --- &    -51.6$^\circ$ &   -50.3$^\circ$ &   -26.8$^\circ$ &   --- &    --- &    -22.7$^\circ$  & 5.0$^\circ$  &  33.3 $^\circ$ \\
Rev 256 HPMONITOR &
2017-011   &    --- &    -51.9$^\circ$ &   -49.6$^\circ$ &   -25.3$^\circ$ &   --- &    --- &    -20.9$^\circ$ &   6.1$^\circ$ &     33.7$^\circ$ \\
Rev 263 HIPHASE &
2017-060   &    --- &    -51.2$^\circ$ &   -50.0$^\circ$ &   -26.0$^\circ$ &  --- &    --- &    --- &    --- &    --- \\
Rev 272 HIPHASE &
2017-123   &    -57.1$^\circ$ &   -52.0$^\circ$ &   -50.2$^\circ$ &   -25.9$^\circ$ &   -22.4$^\circ$ &   -21.5$^\circ$ &   -21.1$^\circ$  & 6.5$^\circ$ &     34.1$^\circ$ \\
Rev 279 HPMONITOR &
2017-170  &     -57.5$^\circ$ &   -53.1$^\circ$ &   -50.5$^\circ$ &   -26.2$^\circ$ &   -22.0$^\circ$ &   -21.3$^\circ$ &   -20.6$^\circ$ &   6.8$^\circ$ &     --- \\
Rev 289 HPMONITOR &
2017-229  &     -57.6$^\circ$ &   -53.7$^\circ$ &   -50.8$^\circ$ &   -26.3$^\circ$ &   -22.1$^\circ$ &   -21.2$^\circ$ & -20.0$^\circ$ &   --- &    --- \\ \hline
\end{tabular}}

$^*$ Location of the slope break that marks the trailing edge of the ML clump
\end{table*}

\begin{table}[tbh]
\caption{Drift rates and accelerations of clump  relative to a co-rotating reference frame with a mean motion of 1751.7$^\circ$/day and an epoch time of 300000000 TDB. Note error bars are based on scatter in the observations given in Table~\ref{d68locs}}
\label{driftfits}
\resizebox{2.75in}{!}{\centerline{\begin{tabular}{cccc}\hline
 Feature & Location$^a$ & Drift rate$^a$ & Acceleration \\ \hline
LL Peak & 32.8$\pm0.1^\circ$ &  1.8$\pm 0.2^\circ$/year & -1.0$\pm0.9^\circ$/year$^2$ \\
L Peak &  3.3$\pm0.2^\circ$  & 4.2$\pm 0.2^\circ$/year & -1.1$\pm0.4^\circ$/year$^2$ \\
M/ML Edge &  -24.4$\pm0.3^\circ$ & 6.1$\pm 0.7^\circ$/year & -4.3$\pm1.5^\circ$/year$^2$ \\
M Peak &  -27.4$\pm0.3^\circ$  & 3.5$\pm 0.1^\circ$/year & -4.4$\pm1.8^\circ$/year$^2$ \\
 T Edge &  -51.3$\pm0.4^\circ$  & 2.4$\pm 0.3^\circ$/year & -2.7$\pm0.7^\circ$/year$^2$ \\
 T Peak &  -53.3$\pm0.6^\circ$  & 2.9$\pm 0.4^\circ$/year & -4.1$\pm1.2^\circ$/year$^2$ \\ \hline
\end{tabular}}}

$^a$ Evaluated on 2016-157
\end{table}

\begin{table*}
\caption{Phase-corrected normal equivalent areas for the clumps, along with the co-rotating longitude ranges used to compute these numbers (and background levels). Note the Rev 263 HIPHASE observation is not included because of its limited longitudinal coverage and because its was observed at exceptionally high phase angles, making the phase corrections more suspect.}
\label{d68ints}
\hspace{-.75in}\resizebox{7.5in}{!}{\begin{tabular}{l c ccccc}\hline
Observation & Date & TN Clump & T Clump & M Clump & L Clump & LL Clump \\ \hline
Rev 211 DRCLOSE &
2015-009 &---- &      ----  &    ----- &     23.8 km$^2$ & ----- \\
Phase =  28$^\circ$ & &  ---- & ---- & ---- & $[-11^\circ,-10^\circ,1^\circ,2^\circ]$ & ---- \\    
Rev 212 DRCLOSE &
 2015-042    & ----  &    9.6 km$^2$   &  ----- & ----- & -----\\     
 Phase =  140$^\circ$ & & ----  &    $[-65^\circ,-64^\circ,-55^\circ,-54^\circ]$  &  ----- & ----- & -----\\  
Rev 231 FNTHPMOV &
2016-033 &   ----- & ----- &   367.4 km$^2$ &   ---- &   ----- \\
 Phase =  127$^\circ$ & & ----  &  ----- &   $[-50^\circ,-49^\circ,-24^\circ,-23^\circ]$   & ----- & -----\\  
Rev 233 FNTHPMOV &
2016-071 &  ---- &     38.2 km$^2$   &   397.8 km$^2$ &  66.4 km$^2$  & 12.2 km$^2$ \\
 Phase =  146$^\circ$ &  &  ---- &  $[-61^\circ,-60^\circ,-50^\circ,-49^\circ]$     &   $[-50^\circ,-49^\circ,-24^\circ,-23^\circ]$ &  $[-6^\circ,-5^\circ,6^\circ,7^\circ]$  & $[27^\circ,28^\circ,36^\circ,37^\circ]$ \\
Rev 245 HPCLOSE &
2016-289 &   ---- &     114.3 km$^2$  &  720.3 km$^2$  &  71.9 km$^2$  & 12.8$^\circ$ \\
 Phase =  156$^\circ$ &  &  ---- &  $[-61^\circ,-60^\circ,-50^\circ,-49^\circ]$     &   $[-50^\circ,-49^\circ,-21^\circ,-20^\circ]$ &  $[-3^\circ,-2^\circ,9^\circ,10^\circ]$  & $[28^\circ,29^\circ,37^\circ,38^\circ]$ \\
Rev 256 HPMONITOR &
2017-011 &     ----- & 150.7 km$^2$ & 648.6 km$^2$ &  75.0 km$^2$ &  8.5 km$^2$ \\
 Phase =  158$^\circ$ &  &  ---- &  $[-61^\circ,-60^\circ,-49^\circ,-48^\circ]$     &   $[-49^\circ,-48^\circ,-20^\circ,-19^\circ]$ &  $[-2^\circ,-1^\circ,11^\circ,12^\circ]$  & $[28^\circ,29^\circ,37^\circ,38^\circ]$ \\

Rev 272 HIPHASE &
2017-123 &     0.5 km$^2$ &  201.7 km$^2$ (201.2 km$^{2})^*$ &  629.8 km$^2$ & 42.8 km$^2$  & 9.3 km$^2$ \\
 Phase =  168$^\circ$ &  &   $[-59^\circ,-58^\circ,-56^\circ,-55^\circ]$  &  $[-61^\circ,-60^\circ,-49^\circ,-48^\circ]$     &   $[-49^\circ,-48^\circ,-20^\circ,-19^\circ]$ &  $[-3^\circ,-2^\circ,12^\circ,13^\circ]$  & $[28^\circ,29^\circ,37^\circ,38^\circ]$ \\
Rev 279 HPMONITOR &
2017-170 &     21.0 km$^2$  & 231.5 km$^2$ (220.0 km$^{2})^*$ & ---- & ---- & ---- \\    
 Phase =  149$^\circ$ &  &   $[-59^\circ,-58^\circ,-56^\circ,-55^\circ]$  &  $[-61^\circ,-60^\circ,-49^\circ,-48^\circ]$     &   
---- & ---- & ---- \\
Rev 289 HPMONITOR &
2017-229 &  120.6 km$^2$ & 331.1 km$^2$ (210.5 km$^{2})^*$ & 699.3 km$^2$ & ---- & ---- \\ 
 Phase =  143$^\circ$ &  &   $[-59^\circ,-58^\circ,-56^\circ,-55^\circ]$  &  $[-61^\circ,-60^\circ,-49^\circ,-48^\circ]$     &   
$[-49^\circ,-48^\circ,-18^\circ,-17^\circ]$ & ---- & ---- \\    \hline
\end{tabular}}
$^*$ Corrected to remove contribution from Clump TN.
\end{table*}

\subsection{Evolution of the clumps' positions and brightnesses}
\label{evolution}

\begin{figure}
\resizebox{3.in}{!}{\includegraphics{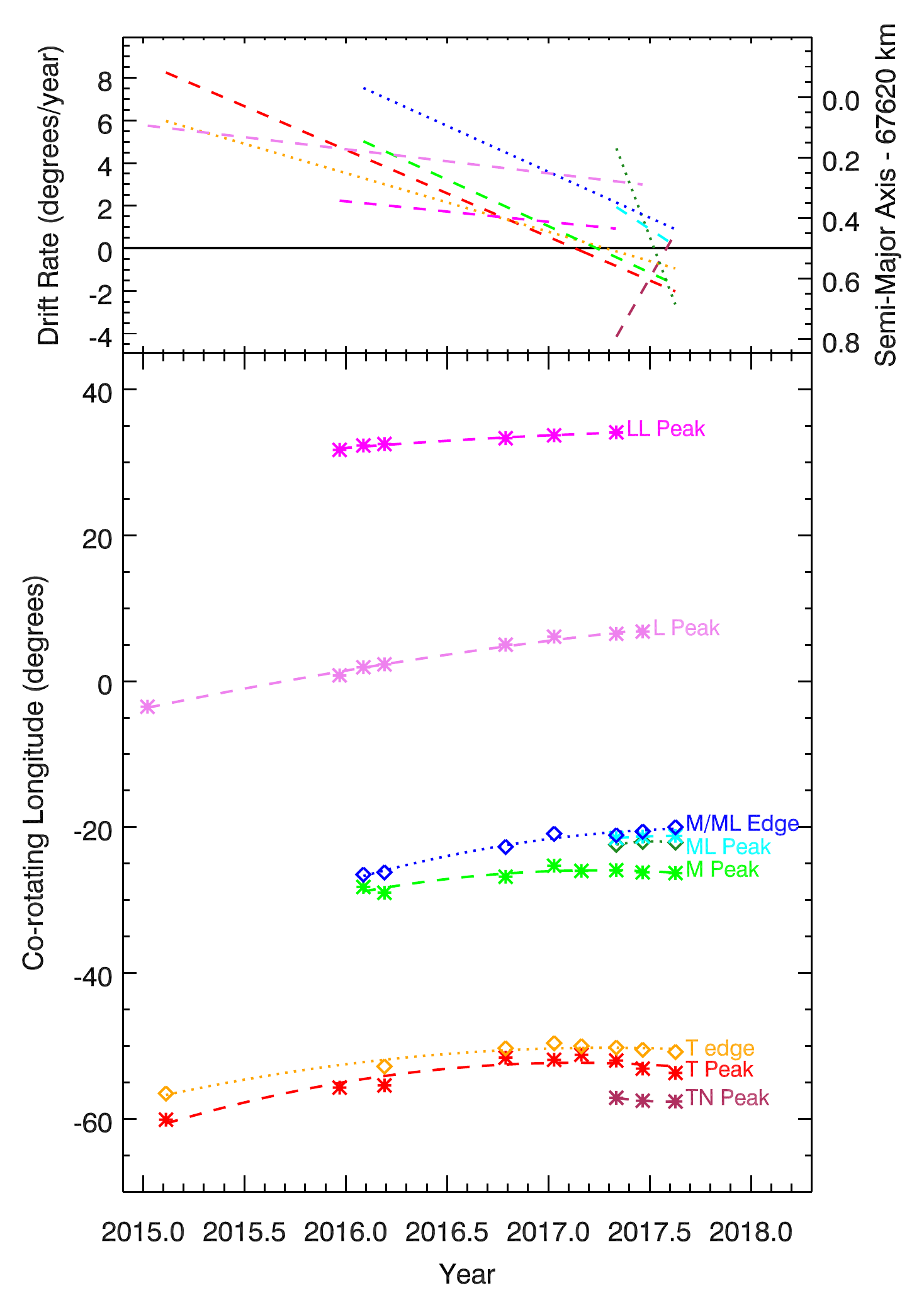}}
\caption{Variations in the locations of the clumps  over time relative to the co-rotating coordinate system. The bottom panel shows the locations of peaks and edges of select clumps over time. The lines show quadratic fits to these data. The top panel shows the drift rates of the clumps relative  to the co-rotating coordinate system moving at 1751.7$^\circ$/day. Also shown are the range of semi-major axes corresponding to these drift rates, computed using the gravitational field harmonics provided in \citet{Durante17}.}
\label{drift}
\end{figure}

\begin{figure}
\resizebox{3in}{!}{\includegraphics{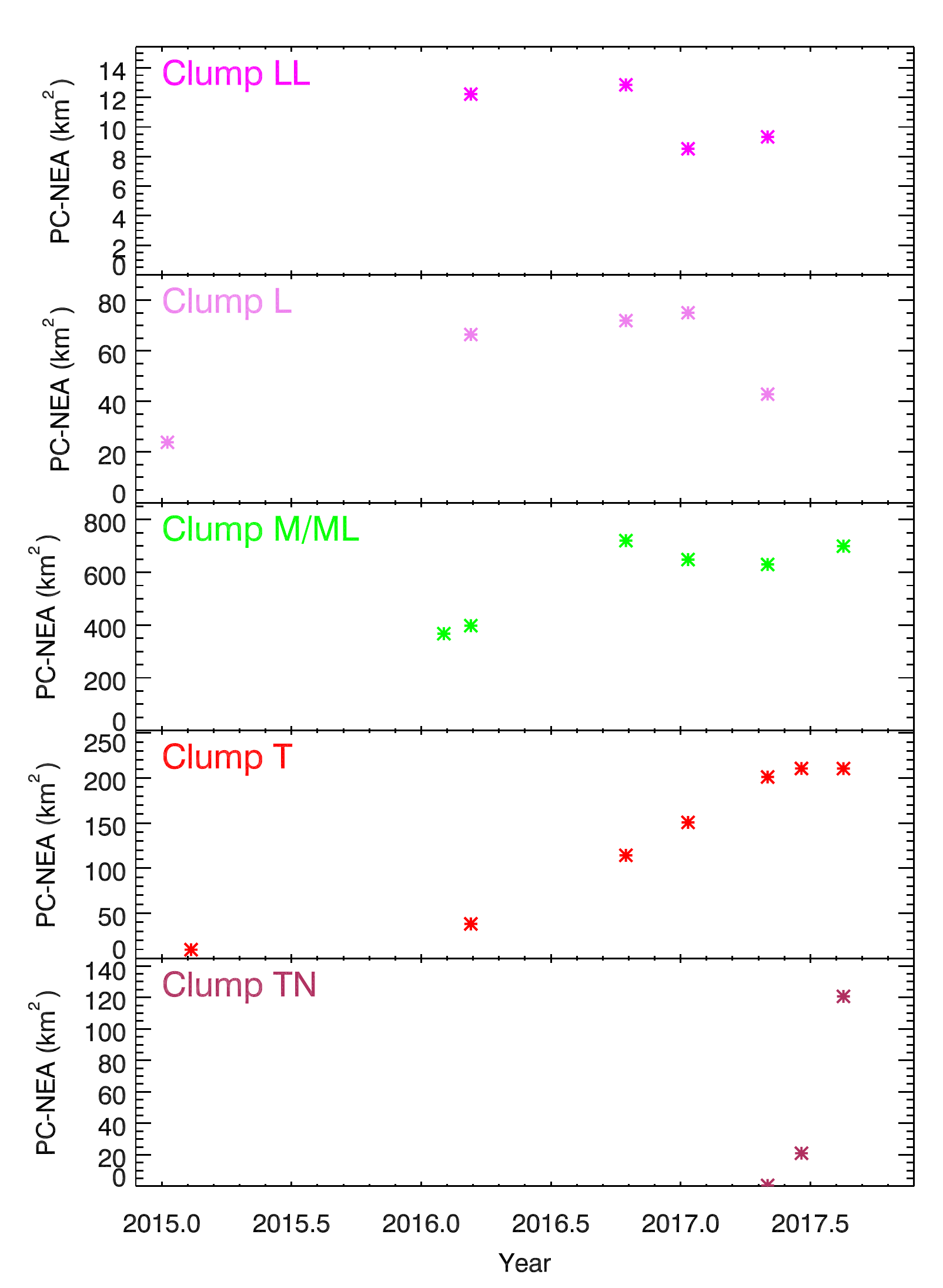}}
\caption{Variations in the integrated brightness of the clumps  over time. Each panel shows the longitudinally-integrated brightness of the various clumps as functions of time. Note that the estimated brightnesses of the T clump have the signal from the TN clump removed.}
\label{bright}
\end{figure}

To better quantify the temporal evolution of these clumps, Tables~\ref{d68locs}-\ref{d68ints} and  Figures~\ref{drift}-\ref{bright} provide summaries of how their positions and integrated brightnesses changed over the course of the Cassini mission. 

Table~\ref{d68locs} provides the observed locations of several clump features. In this table, ``Peak" locations correspond to brightness maxima and ``Edge" locations correspond to minima found just ahead of the T and M/ML clumps (other edges were too indistinct to reliably locate). The highly variable morphology of the clumps made automatic algorithms for locating these features impractical, so the locations given in Table~\ref{d68locs} were instead determined by visual inspection of the profiles. Uncertainties on such numbers cannot be reliably computed {\it a priori}, and so no errors are provided in the table. However, position estimates obtained at roughly the same time differ by only a few tenths of a degree, which suggests that the uncertainties in these parameters are less than half a degree. Figure~\ref{drift} plots these position estimates as functions of time, along with quadratic model fits where the clump is allowed to have a drift rate (relative to the reference rate of 1751.7$^\circ$/day) that varies linearly with time. The parameters derived from these fits (position and drift rate at epoch as well as the acceleration) are provided in Table~\ref{driftfits}, along with uncertainties derived from the scatter in the data points around the trend. 

As shown in the top panel of Figure~\ref{drift}, prior to 2017 all of the clumps were drifting forwards at rates between 2$^\circ$/year and 8$^\circ$/year. However, these drift rates gradually slowed down, with both the M and T clumps beginning to drift backwards at a rate of around $-2^\circ$/year by the end of the Cassini mission. 

Both the average drift rates and the accelerations of these features contain information about the clump's dynamics. For one, the small dispersion in the drift rates associated with these clumps indicate that the clump material is tightly confined in semi-major axis. The dispersion of clump drift rates at any given time is always of order 5$^\circ$/year, which corresponds to a fractional spread in mean motions $\delta n/n \sim 8\times10^{-6}$, which in turn implies a fractional spread in semi-major axes $\delta a/a = 2/3(\delta n/n) \sim 5\times10^{-6}$, or a $\delta a \sim 0.4$ km (see also Figure~\ref{drift}). We can also note that the difference in mean motions between the peak of the M clump and its leading edge is about 2$^\circ$/year (see Table~\ref{driftfits}). If this spreading is due to Keplerian shear, then it implies that the material in this part of the brightest clump has a $\delta a \sim 0.2$ km. The widths of the L and LL clumps are harder to quantify, but we may note that between the Rev 233 FNTHPMOV observation on Day 2016-071 and the Rev 256 HPMONITOR on Day 2017-011 the L clump increased from roughly 5$^\circ$ wide to about 10$^\circ$ wide (see Figure~\ref{fullclump}), which implies the two ends sheared apart at a rate close to 5$^\circ$/year, comparable to the dispersion of the clumps as a whole and the spreading rate of the M clump. All these findings suggest that the clumps consist of material with sub-kilometer spreads in semi-major axis.

Turning to the slow accelerations of the clumps, these correspond to relatively slow {\em changes} in the material's average semi-major axes. For example, the drift rates for the M and T clump changed by roughly 4$^\circ$/year$^2$. This corresponds to an outward radial migration rate of roughly 0.3 km/year (Figure~\ref{drift}). The radial acceleration of the L and LL clumps are smaller, more like 1$^\circ$/year$^2$ (see Table~\ref{driftfits}), but also suggest slow outwards migration. Recall that the mean radius of D68 appears to oscillate with an amplitude of $\sim$ 10 km over a period of roughly 15 years \citep{Hedman14}, which would correspond to maximum radial drift rates of order 3-4 km/year. Note that during this particular time period the ringlet should be moving outward, which is consistent with the observed accelerations of the clumps, but the magnitude of the  radial drift rates are roughly an order of magnitude slower than one would expect for the ringlet as a whole. Hence the connections between the acceleration of the clumps and the overall radial migration of D68 remain obscure.

Next, consider the integrated phase-corrected brightnesses of the clumps, which provide information about the total amount of material within each clump. For each clump in each observation, the total brightness of the clump is computed using the four co-rotating longitudes denoted $[\lambda_1, \lambda_2, \lambda_3, \lambda_4]$ in Table~\ref{d68ints}. First, a linear background is computed based on the average ringlet brightness in two regions on either side of the clump (i.e. between $\lambda_1$ and $\lambda_2$ and between $\lambda_3$ and $\lambda_4$, respectively). Then the observed PC-NEW above this background is integrated over the region between $\lambda_2$ and $\lambda_3$ to determine the phase-corrected normal equivalent area (PC-NEA) of the clump:

\begin{equation}
{\rm PC\mbox{-}NEA}=\int_{\lambda_2}^{\lambda_3} ({\rm PC\mbox{-}NEW}-{\rm PC\mbox{-}NEW}_{back}) r_{D68} d\lambda
\end{equation}
where $r_{D68}=$67,625 km is the radius of the ringlet. In this case, the uncertainties in these numbers are dominated by systematic errors associated with varying image quality and viewing geometry, which are difficult to reliably estimate. Hence, in order to avoid providing a potentially misleading error estimates, no explicit errors on these values are provided here. However, given the scatter in the observed values around mean trends these uncertainties are unlikely to exceed 30\%. 

Table~\ref{d68ints} and Figure~\ref{bright} summarize the total brightness estimates for each of the clumps. Note that for the T clump we also provide the values with the contribution from the superimposed TN clump removed. The clumps show a variety of brightness trends. The L and LL clumps show roughly constant brightness in 2016. Thus these two clumps appear to consist of roughly constant amounts of material that are gradually spreading out over larger and larger longitude ranges, and therefore becoming more indistinct. By contrast, the M, T and TN clumps clearly show initial increases in their total brightness over time. The TN clump's brightening actually accelerated over time, with its PC-NEA going up by 20 km$^2$ in the 47 days between the first two observations, and by another 100 km$^2$ in the following 59 days.  Clump T's increase in brightness could also have accelerated in 2016 if the feature seen in 2015 is really that same clump, but throughout 2016 its brightness increased at a roughly steady rate of 150 km$^2$/year, which is substantially slower than TN's brightening rate. Interestingly, T appears to have stopped brightening around the time TN formed. Clump M's brightening is probably the least well documented, but the early 2016 measurements are roughly one-half the brightness observed in late 2016 and 2017.  Assuming a linear brightness increase between early and mid 2016, this would imply a brightening rate of around 500 km$^2$/year, which is comparable to the brightening rate for the TN clump.

\section{Discussion}
\label{discussion}

Clearly, something happened to D68 in 2014 or 2015 to create the clumps seen at the end of the Cassini Mission. The most obvious explanation for such dramatic and localized increases in brightness is that fine material was released by collisions into or among larger objects located within or nearby D68, similar to the way bright features are thought to form in the F ring \citep{Showalter98, BE02, French12, French14, Murray18}. This scenario will be examined in some detail below.  Section~\ref{sources} examines how much material is needed to produce the visible clumps, and whether suitable source bodies for this material could be lurking in or around D68. Next, Section~\ref{fring} argues that the clumps in D68 are  primarily due to collisions among particles orbiting close to D68, rather than impacts from interplanetary objects. These arguments are based on comparisons between  the properties of the clumps in D68 and those found in the F ring. Finally, Section~\ref{final} examines the spatial and temporal distribution of the clump forming events and what they may imply about the distribution of source bodies within D68.

\subsection{Required sizes of dust sources}
\label{sources}

If the D68 clump material is debris from collisions, then there needs to be objects large enough to produce the observed dust in the vicinity of this ring. How large these source objects would need to be  can be constrained by estimating the total amount of material associated with each clump from the PC-NEA values given in Table~\ref{d68ints}. Since the optical depth of D68 is very low, it is  reasonable to neglect multiple scattering among ring particles. In this case, the equivalent width of the ringlet at a given longitude can be expressed as the following integral over the relevant particle size distribution at a given longitude $\lambda$:

\begin{equation}
EW(\lambda) =\frac{\pi}{|\mu|}\int  N(\lambda, s) Q_{sca}(s)\mathcal{P}(\theta,s)\pi s^2 ds
\end{equation} 
where $\mu$ is the cosine of the emission angle, $Nds$ is the total number of particles per unit length along the ringlet between sizes $s$  and $s+ds$,  $Q_{sca}$ is the particle's scattering coefficient and $\mathcal{P}$ is the particles' phase function. Consistent with the phase function for the ring as a whole given in Equation~\ref{phase},  $\mathcal{P}$ is assumed to integrate to unity over all solid angles, and so the prefactor of $\pi$ is needed to ensure energy conservation. If this expression is integrated over radius and longitude, and then multiplied by $|\mu|$ and divided by the average phase function $P(\theta)$ of the ring from Equation~\ref{phase}, it yields the following expression for the Phase-Corrected Normal Equivalent Area:

\begin{equation}
\begin{split}
{\rm PC\mbox{-}NEA} 
= \pi \int \mathcal{N}(s) Q_{sca}(s)\frac{\mathcal{P}(\theta,s)}{P(\theta)}\pi s^2 ds
\end{split}
\end{equation} 
where  $\mathcal{N}(s) = \int N(\lambda,s) rd\lambda$ is the particle size distribution for the entire clump and $r$ is the mean radius of the ringlet. The total mass of material in this region is given by a similar integral expression over this size distirbution:

\begin{equation}
M =\rho V=\rho \int \mathcal{N}(s) \frac{4}{3}\pi s^3 ds
\end{equation}
where $\rho$ is the mass density of the particles, and $V$ is the total volume of all the particles in the region. These two expressions involve different integrals over the particle size distribution. Still,  there is a relationship between the total mass and the PC-NEA, which can written as:

\begin{equation}
M=\rho \frac{4}{3\pi}({\rm PC\mbox{-}NEA}) \left(s/Q_{\rm sca}\right)_{\rm eff}
\end{equation}
where $\left(s/Q_{\rm sca}\right)_{\rm eff}$ is an effective average particle size/scattering efficiency given by the following ratio of integrals.

\begin{equation}
\left(s/Q_{\rm sca}\right)_{\rm eff}=\frac{\int\mathcal{N}(s) s^3 ds}{\int \mathcal{N}(s) Q_{sca}(s)\frac{\mathcal{P}(\theta,s)}{P(\theta)} s^2 ds}.
\end{equation}
While we still do not have complete knowledge of the particle size distribution, the strongly forward-scattering phase function of the ring suggests that the visible particles are primarily in the  size range of 1-100 microns, and probably have an effective average size of order a few microns \citep{HS15}. We can therefore approximate the total mass of material in the clumps as:

\begin{equation}
\footnotesize
M =40,000\mbox{ kg} \left(\frac{\rho}{1 g/cm^3}\right)\left(\frac{\rm PC-NEA}{100\mbox{ km}^2}\right) \left(\frac{\left(s/Q_{\rm sca}\right)_{\rm eff}}{1\mu m}\right)
\end{equation}
This implies that if $\left(s/Q_{\rm sca}\right)_{\rm eff} \simeq 1 \mu$m, then the LL, L, M, T, and TN clumps would have peak masses of around 7000 kg, 30,000 kg, 300,000 kg, 100,000 kg and  50,000 kg, respectively, and that the total mass of all the visible clumps would be around 500,000 kg.  If gathered into a single object with a density of 1 g/cm$^3$, these objects would all have radii between 1 and 4 meters.\footnote{ Even if we allowed $\left(s/Q_{\rm sca}\right)_{\rm eff}$ to be as large as 10 microns, the M clump would still correspond to a solid object with a radius of around 9 meters.} By way of comparison, for the same assumptions, the mass of the background D68 ringlet would be roughly 5,000,000 kg, or about 10 times larger than the total mass of all of the clumps. Thus the visible clump material is just a small fraction of the dust in the entire ringlet. 

Of course, the above calculation is the {\em minimum} mass required to produce the visible dust clumps, and does not include material released in the form of vapor, particles much smaller than 1 micron in radius or particles much larger than 10 microns in radius.  Still, the above calculations imply that the observed clumps do not require kilometer-scale objects to supply the observed material. Instead the source objects could be comparable in size to the largest typical particles in the C ring \citep{Zebker85, Cuzzi09, Baillie13}. 

The relatively small amount of material required to produce these clumps is also generally compatible with the lack of any direct evidence for any particles larger than the visible dust grains in the available images of D68. Objects  with radii between 1 mm and 1 km can be very difficult to see because their surface-area-to-volume ratios are smaller than small dust grains and  because they are too small to be easily resolved as discrete objects in most images of D68. 

Indirect evidence for such source bodies comes from two high-resolution observations of D68 obtained in 2005 that contained a secondary peak on the inner flank of this ringlet, displaced inwards by 10-20 km from the main D68 ringlet \citep[see Figure 10 of][]{Hedman07}. Unfortunately, no later high-resolution D68 observations obtained during the remainder of the mission covered the same co-rotating longitude range, and no other images showed clear evidence for additional ringlets near D68, so the connections between these secondary peaks and the clumps are rather obscure. However, the two observations where the secondary peaks are clear do appear to have occurred around 0$^\circ$ in the above co-rotating longitude system, and so they could potentially represent material scattered out of D68 by larger objects that later gave rise to the clump material. More detailed analysis would be needed to ascertain whether similar-sized objects could produce both the clumps and the displaced ring material.

Further evidence for a population of larger particles in the vicinity of D68 comes from the in-situ measurements made by the Cassini spacecraft when it passed between the planet and its rings. During this time, the Low Energy Magnetospheric Measurement System (LEMMS) component of the Magnetosphere Imaging Instrument (MIMI) detected a clear reduction in the intensity of protons  and electrons when the spacecraft crossed magnetic field lines that passed near D68 \citep{Krupp18, Kollmann18, Roussos18}. This localized reduction in charged particle flux implies that there is a concentration of material capable of absorbing charged particles around D68. The total mass of this material is still being investigated \citep{Kollmann18}. However, it is worth noting that D68 is the only feature in the D ring interior to D73 that significantly affects the measured plasma densities, despite the fact that D68's brightness relative to its surroundings is not exceptionally high \citep{Hedman07}. Hence it is reasonable to conclude that the reduction in the plasma density around D68 is due to a population of larger particles orbiting in the vicinity of D68, which are invisible to the cameras but efficient absorbers of charged particles.

\subsection{Insights into how the clumps formed from comparisons between D68 and the F ring}
\label{fring}

If the clumps found in D68 are collisional debris, then their closest analogs would be the bright clumps in the F ring, which have also been interpreted as the results of collisions either into or among larger objects within that rings \citep{Showalter98, BE02, Showalter04, Charnoz05, Charnoz09, French12, French14, Murray18}. Indeed, by comparing the overall brightness, motions and evolution of these two different types of clumps, we can gain some insights into what sorts of collisions could be responsible for producing the clumps in D68.

First of all, the clumps in D68 appear to involve much less material than the clumps in the F ring. \citet{French14} provides the most extensive  survey of F-ring clumps to date, which include phase-normalized integrated brightness values. However, some care is needed in comparing the two sets of brightness estimates because \citet{French14} normalized the observed brightness by the ratio of the phase function at the observed phase angle to the phase function at 0$^\circ$ phase. This differs from the normalization used here by a factor of the phase function at 0$^\circ$ phase, which is 0.0095 for the phase function given by Equation~\ref{phase}. Hence the values of PC-NEA need to be multiplied by this factor to obtain ``Phase-Normalized Normal Equivalent Areas''  (PN-NEA) that can be compared with the values given in \citet{French14}. The range of 10-700 km$^2$ in the PC-NEA values for the D68 clumps therefore correspond to PN-NEA values in the range of 0.1-7 km$^2$. By contrast, the F-ring clumps have PN-NEA values ranging between 100 and 20,000 km$^2$, and a few clumps are even brighter than this. Each of the F-ring clumps therefore includes hundreds to thousands of times more material than the D68 clumps. 

The above comparisons imply that the collisions release more material in the F ring than they do in D68. This suggests that the F ring has more abundant and/or larger potential dust sources than D68.
This is a reasonable supposition, since there is abundant evidence from images, occultations and charged-particle data for a population of kilometer-scale moonlets within the F ring \citep{Cuzzi88, Porco05, Murray08, Meinke12, Attree14, Murray18}. By contrast, the generally homogeneous structure of D68 prior to 2015 strongly suggests that such large objects are not common in the vicinity of that ringlet.\footnote{As mentioned in Section~\ref{sources}, there are two high-resolution images that contain evidence for substructure within D68 that could be due to embedded objects. However,  such structures are not seen in any of the more recent images,  which is consistent with such features being relatively rare in this ringlet.} The F ring probably contains more large source bodies because it lies close to Saturn's Roche limit for ice-rich objects, where larger objects can more easily survive and grow, while D68 is located very close to the planet, where tidal forces will inhibit any accumulation of material into larger objects. Hence it is reasonable to expect that there is more source material for clumps in the F ring than in D68. 


Next, consider the range of drift rates and spreading timescales of the clumps in D68 and the F-ring. The clumps in the F ring have drift rates that vary by $\sim 0.2^\circ$/day, or $\sim 100^\circ$/year, and the lengths of individual clumps change at comparable rates \citep{Showalter04, French14, Lam14, Murray18}. These rates are over an order of magnitude larger than the range of drift rates and spreading rates observed in D68. This implies that the material in the F-ring clumps has a much larger spread in semi-major axes than the material in the D68 clumps. Indeed, the spread of drift rates in the F-ring clumps implies that this material spans a semi-major axis range of  order ten kilometers, compared to the sub-kilometer range spanned by D68's clumps. These differences in semi-major axis spreads are also probably responsible for the differences in how long it takes these clumps to fade away. For example, \citet{French12} showed that an exceptionally bright F-ring clump brightened at a roughly constant rate for about 5 months before fading in a quasi-exponential manner with a half-life of roughly 100 days. The fading timescale for this clump is short compared to the brightness evolution timescales of the D68 clumps, whose integrated brightness could remain constant for over a year. This is consistent with the F ring clumps having shorter spreading timescales than the D68 clumps due to the particles' broader semi-major axis range. 

The above differences in the clumps' evolution rates strongly suggests that the material in both these ringlets is primarily released by collisions {\em among} objects within the ring, rather than collisions {\em into} those objects by meteoroids on interplanetary trajectories. If the clumps were created by interplanetary impactors, the velocity dispersion of the debris would be similar for the two rings (in fact, it would probably be higher for D68 because the planet's gravitational focusing would increase the relevant impact speeds), which is clearly not the case. However, if the collisions involve interactions among objects within the ring, then the relative velocities would depend on the velocity dispersion of those objects. While we do not have direct measurements of the orbit parameter dispersion for all the potential source bodies in either the F ring or D68, we may note that while D68 typically appears to be about ten kilometers wide \citep{Hedman07}, the F ring has multiple components that span hundreds of kilometers \citep[see][and references therein]{Murray18}. While the visible material is mostly dust, it is reasonable to expect that the larger particles in the F ring are also more dispersed than the ones in D68, in which case collisions among objects in the F ring happen at higher relative speeds than those within the D68, which could more naturally explain the different range of drift rates and spreading timescales for the two rings. 

Finally, we should note that the extended time it takes for clumps in both D68 and the F ring to reach their maximum brightness is more easily explained if they are both created by collisions among multiple objects within the same ring. A collision involving an interplanetary micrometeoroid would release dust in a short period of time, which is not what is observed either in D68 or in the F ring \citep{French12}. However, if the objects involved in the collision are on nearly the same orbit, then the larger bits of debris from the collision would also be in roughly the same region of phase space, increasing the possibility of repeated collisions, and a gradual release of fine material.

\subsection{Why did the clumps form where and when they did?}
\label{final}

If the above arguments are correct, then the debris seen in the D68 clumps probably arose from collisions among larger objects orbiting close to or within the ringlet. Of course, this immediately raises questions about both the timing of the clump formation and the distribution of the source bodies that gave rise to the clumps. At the moment, there are no clear answers to these questions, but we can at least examine some aspects of these clumps that might be relevant to understanding their origin.

First of all, it is reasonable to ask why clumps only appeared in D68 after 2014. If these clumps are due to collisions among larger objects within the ring, then something must have happened at that time that increased the probability of such collisions. There are two different potential explanations for what could have happened at this time, one involving the internal evolution of D68 itself, and the other involving  impacts by objects from outside the Saturn system.

If one wishes to attribute the timing of clump formation to processes internal to D68, the aspect of this ringlet's structure that is most likely to be relevant is the slow evolution of its mean radius. Prior observations of D68 showed that its mean radius slowly declined by 20 km between 2006 and 2012. Later observations, combined with earlier Voyager images suggest that the mean radius of this ringlet oscillates back and forth with a period of order 15 years \citep{Hedman14}. The visible ringlet's mean radius was therefore moving outwards during 2014-2015. Since the origin of this oscillation is still unclear, the visible dust and larger source bodies could oscillate differently, and so perhaps 2014 corresponded to a critical time where dust was more likely to collide with the source bodies and thereby release additional material. The major problem with such an idea is that in 2014 the visible ringlet was not near either its maximum or its minimum mean radius, and was instead close to the same radius it was in 2011, a time when no obvious clumps appeared. 

If one wants to consider scenarios where something exterior to D68 initiated the clump formation process, then the most likely option would be that one or more interplanetary objects collided with bodies near D68, initiating the collisional cascades that generated the clumps. A challenge for this sort of scenario is that the LL, L, M and T clumps all appear to have formed around the same time, and did not move substantially further apart during the 18 months they were observed. This either means that debris from the original event was able to spread across a region 90$^\circ$ wide, or that multiple impacts struck different parts of D68 around the same time. In practice, the former option appears unlikely, since the observed clumps are well separated and do not appear to be parts of a continuum of debris. The idea that multiple objects could have struck multiple source bodies at about the same time might at first seem equally unlikely. However, there is evidence that Saturn's rings have not only been struck by discrete objects, but also by more extensive debris clouds analogous to meteor storms. In particular, corrugations found in the C and D rings appear to have been generated by such debris fields, which probably represent material released from an object that was torn apart by either tidal forces or a prior impact with the main rings during a previous passage through the Saturn system \citep{Hedman11, Showalter11, Marouf11,  Hedman15, Hedman16}. One could therefore posit that a similar debris cloud passed through D68 in 2014-2015, impacting at least 4 source bodies and so initiating the formation of clumps LL, L, M and T. In principle, such a debris cloud could have also had affects on other parts of Saturn's ring system, but at the moment there is no clear evidence for such a recent event elsewhere in the rings. Hence we cannot yet place strong constraints on exactly what event initiated the formation of D68's clumps. 

Turning to the spatial distribution of the dust sources, the first thing worth noting is that objects large enough to generate the clump debris must have orbits very close to that of D68. Since D68 is a uniquely narrow, isolated ringlet in the otherwise rather broad and smooth inner D ring, this strongly implies that some process is confining material near this ringlet. In principle, the  visible dust could become trapped by a variety of non-gravitational processes, such as resonances with asymmetries in Saturn's electromagnetic field. However, if each clump is debris from collisions involving larger objects, then all those objects would need to have very similar orbits to D68, which strongly suggests that the relevant confining force is not size dependent. The forces responsible for confining D68 are therefore most likely gravitational. Given D68's narrowness, the confinement mechanism most likely involves some sort of resonance with either one of Saturn's moons or some asymmetry in the planet's gravitational field. However, at the moment there is no known resonance with any of Saturn's moons that could explain the observed properties of D68. More generally, the ringlet's lack of strong azimuthal brightness variations prior to 2014, as well as its simple eccentric shape, are not consistent with the orbital perturbations associated with most resonances. 

It is also worth noting that the clumps in D68 do not appear to be randomly distributed. For one, the clumps are all confined to a region roughly 120$^\circ$ across. Furthermore, the TN/T, M/ML, L and LL clumps appear to maintain a suspiciously regular spacing of roughly 30$^\circ$ for the entire time they are observed.  Specifically, the separation between the peaks of the T and M clumps was 26$^\circ\pm1^\circ$, the separation between the M and L clumps was 32$^\circ\pm1^\circ$, and the separation between the  L and LL clumps was 29$^\circ\pm1^\circ$. This suggests that the source bodies are not randomly distributed around D68, but that there is something selecting out particular locations for either the source bodies or the dust release. The lack of strong azimuthal brightness variations in the dust prior to 2014 would be difficult to reconcile with any external force confining dusty material in longitude. Hence it seems more likely that this spacing reflects something about the distribution of the source bodies. Interestingly, the stable solution for roughly 4 equal-mass bodies in the same orbit also has the four objects spanning roughly 120$^\circ$ and being spaced by between 30$^\circ$ and $\sim$40$^\circ$ \citep[][ I thank J. A'Hearn for pointing this out]{Salo88, Renner04}. We may therefore posit that there is some outside force that is trapping material at a particular semi-major axis, which includes both large source bodies and dust. This trapping potential would need to be longitude-independent, allowing the few large bodies in this region have arranged themselves into a stable co-orbital configuration.  At the moment, I am not aware of any phenomenon that can satisfy all these requirements, so more work needs to be done to develop a plausible dynamical explanation for the confinement and structure of D68.

\section{Potential connections to in-situ observations from Cassini's Grand Finale}
\label{prox}

\begin{figure*}
\resizebox{6.5in}{!}{\includegraphics{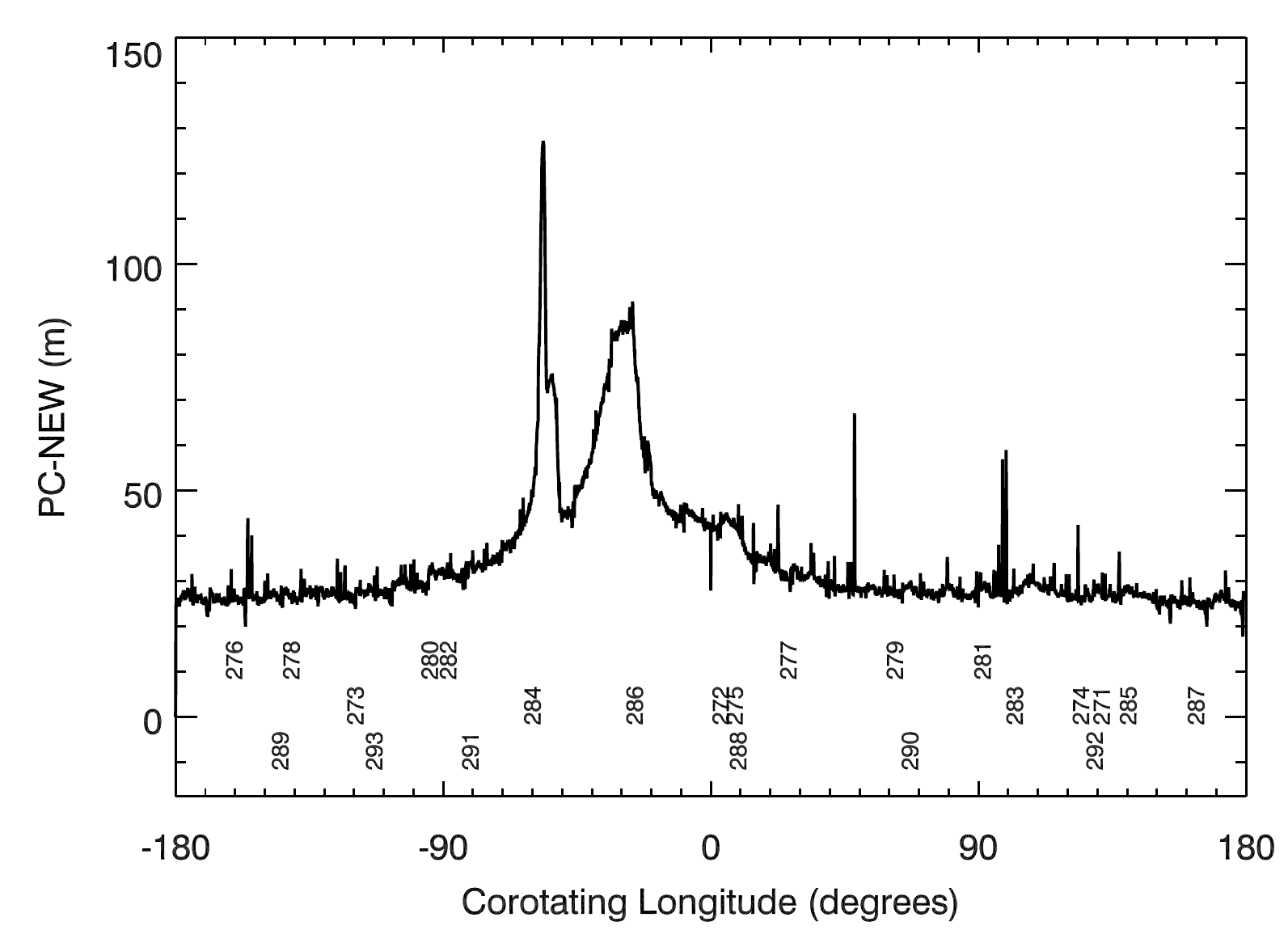}}
\caption{D68 brightness profile derived from the Rev 289 HPMONITOR on day 2017-229. The numbers below the profile mark the longitudes where Cassini passed through the ringplane on each of the Grand Finale orbits. Each number is the designated ``Rev" for each orbit.}
\label{proxcomp}
\end{figure*}

During its Grande Finale, the Cassini spacecraft passed between the planet and D68, enabling it to make in-situ measurements of the material in this region. Some of these measurements revealed azimuthal variations that might be correlated with the D68 clump locations \citep[e.g.][]{Waite18}. While the visible material in D68 appears to be strongly confined in semi-major axis, the clump-forming events could potentially release smaller particles and molecules that could more easily reach the spacecraft. In addition, the event that triggered clump formation in D68 could have had larger-scale effects on the planet and/or its rings that might have influenced these measurements. Hence, for the sake of completeness, Figure~\ref{proxcomp} shows where the Cassini spacecraft passed relative to these clumps on all of the relevant orbits. The spacecraft clearly sampled a wide range of longitudes relative to the clumps during its final orbits, allowing a variety of hypotheses to be tested regarding connections between D68's clumps and the in-situ measurements.

\section{Summary and Conclusions}
\label{summary}

The main results of the above analysis of D68's longitudinal structure are the following:

\bigskip

\bigskip

\begin{itemize}
\item Sometime in 2014 or 2015 a series of four bright clumps (here designated T, M, L and LL) appeared in D68. \newline

\item The material in two of these clumps (L and LL) slowly spread over time, making the clumps less distinct.

\item The two other clumps (T and M) became progressively brighter over the course of 2016, and appeared to give rise to additional structures in 2017.  \newline

\item The spreading rates and dispersion in drift velocities suggest that the material in all these clumps spans less than a kilometer in semi-major axis.

\item The total amount of material visible in the clumps could come from a few objects less than 10 meters in radius. 

\item These clumps could have been produced by collisions among larger objects orbiting within or very close to D68.

\item The spatial distribution of these clumps may provide new insights into how material is confined around D68.
\end{itemize}

\section*{Acknowledgements}

The author thanks NASA, the Cassini project and especially the imaging team for obtaining the data used in this work. He also thanks J. A'Hearn, R. Chancia, M.R. Showalter, D.P. Hamilton and P.D. Nicholson for helpful conversations about various topics related to this ringlet, as well as the two reviewers whose comments improved this manuscript. This work was supported in part by Cassini Data Analysis and Participating Scientist Program grant NNX15AQ67G.


\begin{thebibliography}{}

\bibitem[{Attree} {\em et~al.}(2014){Attree}, {Murray}, {Williams}, and
  {Cooper}]{Attree14}
{Attree}, N.~O., C.~D. {Murray}, G.~A. {Williams},\ and N.~J. {Cooper} 2014.
\newblock {A survey of low-velocity collisional features in Saturn's F ring}.
\newblock {\em Icarus\/}~{\bf 227}, 56--66.

\bibitem[{Bailli{\'e}} {\em et~al.}(2013){Bailli{\'e}}, {Colwell}, {Esposito},
  and {Lewis}]{Baillie13}
{Bailli{\'e}}, K., J.~E. {Colwell}, L.~W. {Esposito},\ and M.~C. {Lewis} 2013.
\newblock {Meter-sized Moonlet Population in Saturn's C Ring and Cassini
  Division}.
\newblock {\em AJ\/}~{\bf 145}, 171.

\bibitem[{Barbara} and {Esposito}(2002){Barbara} and {Esposito}]{BE02}
{Barbara}, J.~M.,\ and L.~W. {Esposito} 2002.
\newblock {Moonlet Collisions and the Effects of Tidally Modified Accretion in
  Saturn's F Ring}.
\newblock {\em Icarus\/}~{\bf 160}, 161--171.

\bibitem[{Charnoz}(2009){Charnoz}]{Charnoz09}
{Charnoz}, S. 2009.
\newblock {Physical collisions of moonlets and clumps with the Saturn's F-ring
  core}.
\newblock {\em Icarus\/}~{\bf 201}, 191--197.

\bibitem[{Charnoz} {\em et~al.}(2005){Charnoz}, {Porco}, {D{\'e}au}, {Brahic},
  {Spitale}, {Bacques}, and {Baillie}]{Charnoz05}
{Charnoz}, S., C.~C. {Porco}, E.~{D{\'e}au}, A.~{Brahic}, J.~N. {Spitale},
  G.~{Bacques},\ and K.~{Baillie} 2005.
\newblock {Cassini discovers a kinematic spiral ring around Saturn}.
\newblock {\em Science\/}~{\bf 310}, 1300--1304.

\bibitem[{Cuzzi} {\em et~al.}(2009){Cuzzi}, {Clark}, {Filacchione}, {French},
  {Johnson}, {Marouf}, and {Spilker}]{Cuzzi09}
{Cuzzi}, J., R.~{Clark}, G.~{Filacchione}, R.~{French}, R.~{Johnson},
  E.~{Marouf},\ and L.~{Spilker} 2009.
\newblock {\em {Ring Particle Composition and Size Distribution}}, pp.\  459.

\bibitem[{Cuzzi} and {Burns}(1988){Cuzzi} and {Burns}]{Cuzzi88}
{Cuzzi}, J.~N.,\ and J.~A. {Burns} 1988.
\newblock {Charged particle depletion surrounding Saturn's F ring - Evidence
  for a moonlet belt?}
\newblock {\em Icarus\/}~{\bf 74}, 284--324.

\bibitem[{Durante}(2017){Durante}]{Durante17}
{Durante}, D. 2017.
\newblock {\em {The gravity fields of Jupiter and Saturn as determined by Juno
  and Cassini}}.
\newblock Ph.\ D. thesis, Sapienza Universita de Roma.

\bibitem[{Ferrari} and {Brahic}(1997){Ferrari} and {Brahic}]{FB97}
{Ferrari}, C.,\ and A.~{Brahic} 1997.
\newblock {Arcs and clumps in the Encke division of Saturn's rings}.
\newblock {\em \planss\/}~{\bf 45}, 1051--1067.

\bibitem[{French} {\em et~al.}(1991){French}, {Nicholson}, {Porco}, and
  {Marouf}]{French91}
{French}, R.~G., P.~D. {Nicholson}, C.~C. {Porco},\ and E.~A. {Marouf} 1991.
\newblock {Dynamics and structure of the Uranian rings}.
\newblock In J.~T. {Bergstralh}, E.~D. {Miner}, and M.~S. {Matthews} (Eds.),
  {\em Uranus}, pp.\  327--409.

\bibitem[{French} {\em et~al.}(2014){French}, {Hicks}, {Showalter}, {Antonsen},
  and {Packard}]{French14}
{French}, R.~S., S.~K. {Hicks}, M.~R. {Showalter}, A.~K. {Antonsen},\ and D.~R.
  {Packard} 2014.
\newblock {Analysis of clumps in Saturn's F ring from Voyager and Cassini}.
\newblock {\em Icarus\/}~{\bf 241}, 200--220.

\bibitem[{French} {\em et~al.}(2012){French}, {Showalter}, {Sfair},
  {Arg{\"u}elles}, {Pajuelo}, {Becerra}, {Hedman}, and {Nicholson}]{French12}
{French}, R.~S., M.~R. {Showalter}, R.~{Sfair}, C.~A. {Arg{\"u}elles},
  M.~{Pajuelo}, P.~{Becerra}, M.~M. {Hedman},\ and P.~D. {Nicholson} 2012.
\newblock {The brightening of Saturn's F ring}.
\newblock {\em Icarus\/}~{\bf 219}, 181--193.

\bibitem[{Hedman} {\em et~al.}(2011){Hedman}, {Burns}, {Evans}, {Tiscareno},
  and {Porco}]{Hedman11}
{Hedman}, M.~M., J.~A. {Burns}, M.~W. {Evans}, M.~S. {Tiscareno},\ and C.~C.
  {Porco} 2011.
\newblock {Saturn's curiously corrugated C ring}.
\newblock {\em Science\/}~{\bf 332}, 708--712.

\bibitem[{Hedman} {\em et~al.}(2013){Hedman}, {Burns}, {Hamilton}, and
  {Showalter}]{Hedman13}
{Hedman}, M.~M., J.~A. {Burns}, D.~P. {Hamilton},\ and M.~R. {Showalter} 2013.
\newblock {Of horseshoes and heliotropes: Dynamics of dust in the Encke Gap}.
\newblock {\em Icarus\/}~{\bf 223}, 252--276.

\bibitem[{Hedman} {\em et~al.}(2015){Hedman}, {Burns}, and
  {Showalter}]{Hedman15}
{Hedman}, M.~M., J.~A. {Burns},\ and M.~R. {Showalter} 2015.
\newblock {Corrugations and eccentric spirals in Saturn's D ring: New insights
  into what happened at Saturn in 1983}.
\newblock {\em Icarus\/}~{\bf 248}, 137--161.

\bibitem[{Hedman} {\em et~al.}(2007a){Hedman}, {Burns}, {Showalter}, {Porco},
  {Nicholson}, {Bosh}, {Tiscareno}, {Brown}, {Buratti}, {Baines}, and
  {Clark}]{Hedman07}
{Hedman}, M.~M., J.~A. {Burns}, M.~R. {Showalter}, C.~C. {Porco}, P.~D.
  {Nicholson}, A.~S. {Bosh}, M.~S. {Tiscareno}, R.~H. {Brown}, B.~J. {Buratti},
  K.~H. {Baines},\ and R.~{Clark} 2007a.
\newblock {Saturn's dynamic D ring}.
\newblock {\em Icarus\/}~{\bf 188}, 89--107.

\bibitem[{Hedman} {\em et~al.}(2007b){Hedman}, {Burns}, {Tiscareno}, {Porco},
  {Jones}, {Roussos}, {Krupp}, {Paranicas}, and {Kempf}]{Hedman07g}
{Hedman}, M.~M., J.~A. {Burns}, M.~S. {Tiscareno}, C.~C. {Porco}, G.~H.
  {Jones}, E.~{Roussos}, N.~{Krupp}, C.~{Paranicas},\ and S.~{Kempf} 2007b.
\newblock {The source of Saturn's G ring}.
\newblock {\em Science\/}~{\bf 317}, 653--657.

\bibitem[{Hedman} {\em et~al.}(2014){Hedman}, {Burt}, {Burns}, and
  {Showalter}]{Hedman14}
{Hedman}, M.~M., J.~A. {Burt}, J.~A. {Burns},\ and M.~R. {Showalter} 2014.
\newblock {Non-circular features in Saturn's D ring: D68}.
\newblock {\em Icarus\/}~{\bf 233}, 147--162.

\bibitem[{Hedman} {\em et~al.}(2009){Hedman}, {Murray}, {Cooper}, {Tiscareno},
  {Beurle}, {Evans}, and {Burns}]{Hedman09}
{Hedman}, M.~M., C.~D. {Murray}, N.~J. {Cooper}, M.~S. {Tiscareno},
  K.~{Beurle}, M.~W. {Evans},\ and J.~A. {Burns} 2009.
\newblock {Three tenuous rings/arcs for three tiny moons}.
\newblock {\em Icarus\/}~{\bf 199}, 378--386.

\bibitem[{Hedman} and {Showalter}(2016){Hedman} and {Showalter}]{Hedman16}
{Hedman}, M.~M.,\ and M.~R. {Showalter} 2016.
\newblock {A new pattern in Saturn's D ring created in late 2011}.
\newblock {\em Icarus\/}~{\bf 279}, 155--165.

\bibitem[{Hedman} and {Stark}(2015){Hedman} and {Stark}]{HS15}
{Hedman}, M.~M.,\ and C.~C. {Stark} 2015.
\newblock {Saturn's G and D Rings Provide Nearly Complete Measured Scattering
  Phase Functions of Nearby Debris Disks}.
\newblock {\em ApJ\/}~{\bf 811}, 67.

\bibitem[{Hubbard} {\em et~al.}(1986){Hubbard}, {Brahic}, {Sicardy}, {Elicer},
  {Roques}, and {Vilas}]{Hubbard86}
{Hubbard}, W.~B., A.~{Brahic}, B.~{Sicardy}, L.-R. {Elicer}, F.~{Roques},\ and
  F.~{Vilas} 1986.
\newblock {Occultation detection of a Neptunian ring-like arc}.
\newblock {\em \nat\/}~{\bf 319}, 636--640.

\bibitem[{Kollmann} {\em et~al.}(2018){Kollmann}, {Roussos}, {Kotova},
  {Regoli}, {Mitchell}, {Carbary}, {Clark}, {Krupp}, and
  {Paranicas}]{Kollmann18}
{Kollmann}, P., E.~{Roussos}, A.~{Kotova}, L.~{Regoli}, D.~G. {Mitchell},
  J.~{Carbary}, G.~{Clark}, N.~{Krupp},\ and C.~{Paranicas} 2018.
\newblock {Saturn's Innermost Radiation Belt Throughout and Inward of the
  D-Ring}.
\newblock {\em \grl\/}~{\bf 45}, 10.

\bibitem[{Krupp} {\em et~al.}(2018){Krupp}, {Roussos}, {Kollmann}, {Mitchell},
  {Paranicas}, {Krimigis}, {Hamilton}, {Hedman}, and {Dougherty}]{Krupp18}
{Krupp}, N., E.~{Roussos}, P.~{Kollmann}, D.~G. {Mitchell}, C.~P. {Paranicas},
  S.~M. {Krimigis}, D.~C. {Hamilton}, M.~{Hedman},\ and M.~K. {Dougherty} 2018.
\newblock {Energetic Neutral and Charged Particle Measurements in the Inner
  Saturnian Magnetosphere During the Grand Finale Orbits of Cassini 2016/2017}.
\newblock {\em \grl\/}~{\bf 45}, 10.

\bibitem[{Lam}(2014){Lam}]{Lam14}
{Lam}, W.~F. 2014.
\newblock {\em {Clumping features in Saturn's F ring}}.
\newblock M. Phil. Thesis for QMUL.

\bibitem[{Marouf} {\em et~al.}(2011){Marouf}, {French}, {Rappaport}, {Wong},
  {McGhee-French}, and {Anabtawi}]{Marouf11}
{Marouf}, E., R.~{French}, N.~{Rappaport}, K.~{Wong}, C.~{McGhee-French},\ and
  A.~{Anabtawi} 2011.
\newblock {Uncovering of Small-Scale Quasi-Periodic Structure in Saturn's Ring
  C and Possible Origin}.
\newblock In {\em EPSC-DPS Joint Meeting 2011}, pp.\  265.

\bibitem[{Meinke} {\em et~al.}(2012){Meinke}, {Esposito}, {Albers}, and
  {Srem{\v c}evi{\'c}}]{Meinke12}
{Meinke}, B.~K., L.~W. {Esposito}, N.~{Albers},\ and M.~{Srem{\v c}evi{\'c}}
  2012.
\newblock {Classification of F ring features observed in Cassini UVIS
  occultations}.
\newblock {\em Icarus\/}~{\bf 218}, 545--554.

\bibitem[{Murray} {\em et~al.}(2008){Murray}, {Beurle}, {Cooper}, {Evans},
  {Williams}, and {Charnoz}]{Murray08}
{Murray}, C.~D., K.~{Beurle}, N.~J. {Cooper}, M.~W. {Evans}, G.~A. {Williams},\
  and S.~{Charnoz} 2008.
\newblock {The determination of the structure of Saturn's F ring by nearby
  moonlets}.
\newblock {\em Nature\/}~{\bf 453}, 739--744.

\bibitem[{Murray} and {French}(2018){Murray} and {French}]{Murray18}
{Murray}, C.~D.,\ and R.~S. {French} 2018.
\newblock {The F Ring of Saturn}.
\newblock In {M.S. Tiscareno and C.D. Murray} (Ed.), {\em Planetary Ring
  Systems}, pp.\  338--362. Cambridge U. Press.

\bibitem[{Namouni} and {Porco}(2002){Namouni} and {Porco}]{NP02}
{Namouni}, F.,\ and C.~{Porco} 2002.
\newblock {The confinement of Neptune's ring arcs by the moon Galatea}.
\newblock {\em \nat\/}~{\bf 417}, 45--47.

\bibitem[{Porco}(1991){Porco}]{Porco91}
{Porco}, C.~C. 1991.
\newblock {An explanation for Neptune's ring arcs}.
\newblock {\em Science\/}~{\bf 253}, 995--1001.

\bibitem[{Porco} {\em et~al.}(2005){Porco}, {Baker}, {Barbara}, {Beurle},
  {Brahic}, {Burns}, {Charnoz}, {Cooper}, {Dawson}, {Del Genio}, {Denk},
  {Dones}, {Dyudina}, {Evans}, {Giese}, {Grazier}, {Helfenstein}, {Ingersoll},
  {Jacobson}, {Johnson}, {McEwen}, {Murray}, {Neukum}, {Owen}, {Perry},
  {Roatsch}, {Spitale}, {Squyres}, {Thomas}, {Tiscareno}, {Turtle}, {Vasavada},
  {Veverka}, {Wagner}, and {West}]{Porco05}
{Porco}, C.~C., E.~{Baker}, J.~{Barbara}, K.~{Beurle}, A.~{Brahic}, J.~A.
  {Burns}, S.~{Charnoz}, N.~{Cooper}, D.~D. {Dawson}, A.~D. {Del Genio},
  T.~{Denk}, L.~{Dones}, U.~{Dyudina}, M.~W. {Evans}, B.~{Giese}, K.~{Grazier},
  P.~{Helfenstein}, A.~P. {Ingersoll}, R.~A. {Jacobson}, T.~V. {Johnson},
  A.~{McEwen}, C.~D. {Murray}, G.~{Neukum}, W.~M. {Owen}, J.~{Perry},
  T.~{Roatsch}, J.~{Spitale}, S.~{Squyres}, P.~{Thomas}, M.~{Tiscareno},
  E.~{Turtle}, A.~R. {Vasavada}, J.~{Veverka}, R.~{Wagner},\ and R.~{West}
  2005.
\newblock {Cassini Imaging Science: Initial Results on Saturn's Rings and Small
  Satellites}.
\newblock {\em Science\/}~{\bf 307}, 1226--1236.

\bibitem[{Porco} {\em et~al.}(2004){Porco}, {West}, {Squyres}, {McEwen},
  {Thomas}, {Murray}, {Del Genio}, {Ingersoll}, {Johnson}, {Neukum}, {Veverka},
  {Dones}, {Brahic}, {Burns}, {Haemmerle}, {Knowles}, {Dawson}, {Roatsch},
  {Beurle}, and {Owen}]{Porco04}
{Porco}, C.~C., R.~A. {West}, S.~{Squyres}, A.~{McEwen}, P.~{Thomas}, C.~D.
  {Murray}, A.~{Del Genio}, A.~P. {Ingersoll}, T.~V. {Johnson}, G.~{Neukum},
  J.~{Veverka}, L.~{Dones}, A.~{Brahic}, J.~A. {Burns}, V.~{Haemmerle},
  B.~{Knowles}, D.~{Dawson}, T.~{Roatsch}, K.~{Beurle},\ and W.~{Owen} 2004.
\newblock {Cassini imaging science: Instrument characteristics and anticipated
  scientific investigations at Saturn}.
\newblock {\em \ssr\/}~{\bf 115}, 363--497.

\bibitem[{Renner} and {Sicardy}(2004){Renner} and {Sicardy}]{Renner04}
{Renner}, S.,\ and B.~{Sicardy} 2004.
\newblock {Stationary Configurations for Co-orbital Satellites with Small
  Arbitrary Masses}.
\newblock {\em Celestial Mechanics and Dynamical Astronomy\/}~{\bf 88},
  397--414.

\bibitem[{Renner} {\em et~al.}(2014){Renner}, {Sicardy}, {Souami}, {Carry}, and
  {Dumas}]{Renner14}
{Renner}, S., B.~{Sicardy}, D.~{Souami}, B.~{Carry},\ and C.~{Dumas} 2014.
\newblock {Neptune's ring arcs: VLT/NACO near-infrared observations and a model
  to explain their stability}.
\newblock {\em \aap\/}~{\bf 563}, A133.

\bibitem[{Roussos} {\em et~al.}(2018){Roussos}, {Kollmann}, {Krupp}, {Kotova},
  {Regoli}, {Paranicas}, {Mitchell}, {Krimigis}, {Hamilton}, {Brandt},
  {Carbary}, {Christon}, {Dialynas}, {Dandouras}, {Hill}, {Ip}, {Jones},
  {Livi}, {Mauk}, {Palmaerts}, {Roelof}, {Rymer}, {Sergis}, and
  {Smith}]{Roussos18}
{Roussos}, E., P.~{Kollmann}, N.~{Krupp}, A.~{Kotova}, L.~{Regoli},
  C.~{Paranicas}, D.~G. {Mitchell}, S.~M. {Krimigis}, D.~{Hamilton},
  P.~{Brandt}, J.~{Carbary}, S.~{Christon}, K.~{Dialynas}, I.~{Dandouras},
  M.~E. {Hill}, W.~H. {Ip}, G.~H. {Jones}, S.~{Livi}, B.~H. {Mauk},
  B.~{Palmaerts}, E.~C. {Roelof}, A.~{Rymer}, N.~{Sergis},\ and H.~T. {Smith}
  2018.
\newblock {A radiation belt of energetic protons located between Saturn and its
  rings}.
\newblock {\em Science\/}~{\bf 362}, aat1962.

\bibitem[{Salo} and {Yoder}(1988){Salo} and {Yoder}]{Salo88}
{Salo}, H.,\ and C.~F. {Yoder} 1988.
\newblock {The dynamics of coorbital satellite systems}.
\newblock {\em A\&A\/}~{\bf 205}, 309--327.

\bibitem[{Showalter} {\em et~al.}(2017){Showalter}, {Lissauer}, {de Pater}, and
  {French}]{Showalter17}
{Showalter}, M., J.~J. {Lissauer}, I.~{de Pater},\ and R.~S. {French} 2017.
\newblock {A Three-Body Resonance Confines the Ring-Arcs of Neptune}.
\newblock In {\em AAS/Division for Planetary Sciences Meeting Abstracts \#49},
  Volume~49 of {\em AAS/Division for Planetary Sciences Meeting Abstracts},
  pp.\  104.01.

\bibitem[{Showalter}(1996){Showalter}]{Showalter96}
{Showalter}, M.~R. 1996.
\newblock {Saturn's D ring in the Voyager images}.
\newblock {\em Icarus\/}~{\bf 124}, 677--689.

\bibitem[{Showalter}(1998){Showalter}]{Showalter98}
{Showalter}, M.~R. 1998.
\newblock {Detection of Centimeter-Sized Meteoroid Impact Events in Saturn's F
  Ring}.
\newblock {\em Science\/}~{\bf 282}, 1099.

\bibitem[{Showalter}(2004){Showalter}]{Showalter04}
{Showalter}, M.~R. 2004.
\newblock {Disentangling Saturn's F Ring. I. Clump orbits and lifetimes}.
\newblock {\em Icarus\/}~{\bf 171}, 356--371.

\bibitem[{Showalter} {\em et~al.}(2011){Showalter}, {Hedman}, and
  {Burns}]{Showalter11}
{Showalter}, M.~R., M.~M. {Hedman},\ and J.~A. {Burns} 2011.
\newblock {The impact of Comet Shoemaker-Levy 9 sends ripples through the rings
  of Jupiter}.
\newblock {\em Science\/}~{\bf 332}, 711--.

\bibitem[{Sicardy} {\em et~al.}(1991){Sicardy}, {Roques}, and
  {Brahic}]{Sicardy91}
{Sicardy}, B., F.~{Roques},\ and A.~{Brahic} 1991.
\newblock {Neptune's rings, 1983-1989: Ground-based stellar occultation
  observations. I - Ring-like arc detections}.
\newblock {\em \icarus\/}~{\bf 89}, 220--243.

\bibitem[{Waite} {\em et~al.}(2018){Waite}, {Perryman}, {Perry}, {Miller},
  {Bell}, {Cravens}, {Glein}, {Grimes}, {Hedman}, {Cuzzi}, {Brockwell},
  {Teolis}, {Moore}, {Mitchell}, {Persoon}, {Kurth}, {Wahlund}, {Morooka},
  {Hadid}, {Chocron}, {Walker}, {Nagy}, {Yelle}, {Ledvina}, {Johnson}, {Tseng},
  {Tucker}, and {Ip}]{Waite18}
{Waite}, J.~H., R.~S. {Perryman}, M.~E. {Perry}, K.~E. {Miller}, J.~{Bell},
  T.~E. {Cravens}, C.~R. {Glein}, J.~{Grimes}, M.~{Hedman}, J.~{Cuzzi},
  T.~{Brockwell}, B.~{Teolis}, L.~{Moore}, D.~G. {Mitchell}, A.~{Persoon},
  W.~S. {Kurth}, J.-E. {Wahlund}, M.~{Morooka}, L.~Z. {Hadid}, S.~{Chocron},
  J.~{Walker}, A.~{Nagy}, R.~{Yelle}, S.~{Ledvina}, R.~{Johnson}, W.~{Tseng},
  O.~J. {Tucker},\ and W.-H. {Ip} 2018.
\newblock {Chemical interactions between Saturn's atmosphere and its rings}.
\newblock {\em Science\/}~{\bf 362}, aat2382.

\bibitem[{West} {\em et~al.}(2010){West}, {Knowles}, {Birath}, {Charnoz}, {Di
  Nino}, {Hedman}, {Helfenstein}, {McEwen}, {Perry}, {Porco}, {Salmon},
  {Throop}, and {Wilson}]{West10}
{West}, R., B.~{Knowles}, E.~{Birath}, S.~{Charnoz}, D.~{Di Nino}, M.~{Hedman},
  P.~{Helfenstein}, A.~{McEwen}, J.~{Perry}, C.~{Porco}, J.~{Salmon},
  H.~{Throop},\ and D.~{Wilson} 2010.
\newblock {In-flight calibration of the Cassini imaging science sub-system
  cameras}.
\newblock {\em \planss\/}~{\bf 58}, 1475--1488.

\bibitem[{Zebker} {\em et~al.}(1985){Zebker}, {Marouf}, and {Tyler}]{Zebker85}
{Zebker}, H.~A., E.~A. {Marouf},\ and G.~L. {Tyler} 1985.
\newblock {Saturn's rings - Particle size distributions for thin layer model}.
\newblock {\em Icarus\/}~{\bf 64}, 531--548.

\end{thebibliography}

\end{document}